\documentclass[onecolumn]{elsart3p}
\usepackage{graphicx,amssymb}
\usepackage{amsmath,psfrag}
\journal{Computer Physics Communications} 
\begin{document}
\newcommand{\mb}{\mathbf}
\begin{frontmatter}

\title{Wien2wannier: From linearized augmented plane waves to maximally localized Wannier functions} 

\author[Augsburg,Prague]{Jan Kune\v{s}}
\corauth[cor]{Corresponding author.}
\ead{kunes@fzu.cz},
\author[Tokyo,Tokyo2]{Ryotaro Arita},
\author[Vienna]{Philipp Wissgott},
\author[Vienna]{Alessandro Toschi},
\author[Kyoto]{Hiroaki Ikeda}
\author[Vienna]{Karsten Held}

\address[Augsburg]{Theoretical Physics III, Center for Electronic Correlations
and Magnetism, Institute of Physics, University of Augsburg, Augsburg 86135, Germany}
\address[Prague]{Institute of Physics, Academy of Sciences of the Czech Republic,
Cukrovarnick\'{a} 10, CZ-162 53 Prague 6, Czech Republic}
\address[Tokyo]{CREST, JST, and Department of Applied Physics, University of Tokyo, Hongo, Tokyo 113-8656, Japan}
\address[Tokyo2]{Department of Applied Physics, University of Tokyo,
Hongo, Tokyo 113-8656, Japan}
\address[Vienna]{Institute for Solid State Physics, Vienna University of Technology, Vienna, Austria}
\address[Kyoto]{Department of Physics, Kyoto University, Kyoto, 606-8502, Japan}

\begin{abstract}

We present an implementaion of interface between the full-potential linearized augmented plane wave package Wien2k and the wannier90 code for the construction of  maximally localized Wannier functions. 
The FORTRAN code and a documentation is made available and results 
are discussed for SrVO$_3$, Sr$_2$IrO$_4$ (including spin-orbit coupling),
LaFeAsO, and FeSb$_2$.
\end{abstract}

 
\end{frontmatter} 
 
\section{Introduction} 
 

In recent years, we have seen a revival of Wannier's idea \cite{Wannier37} 
to express the single electron excitations in a periodic potential
in terms of ``orthogonal `atomic' wave functions'',
later coined Wannier functions. 
While the calculation of bandstructures as, e.g., 
within  density functional theory (DFT) \cite{Hohenberg64a}
is most efficiently performed in the basis of extended wave functions indexed by reciprocal-space vectors ${\bf k}$,
the complementary Wannier picture is often useful - if not essential.
The latter does not only provide insight into the nature of
 chemical bonding, but the localized Wannier 
orbitals are more suitable to
describe physical phenomena where the local aspect is important.
These range from the dielectric polarization \cite{Vanderbilt93}, e.g.,
in the context of ferroelectrics \cite{Marzari98a,Tokura06},
and orbital polarization \cite{orbitalPol} to
molecular dynamics \cite{MD} and transport through nano structures
\cite{Nanotransport1,Nanotransport2}.

Also to describe electronic correlations beyond
the local density approximation (LDA) to DFT, Wannier-like orbitals
are very helpful. 
Here they allow us to limit the interactions to a 
computationally manageable subspace, such as
the local interactions between electrons on the same
transition-metal atom in the $d$-orbitals. The on-site term represents
not only the largest constribution to the interaction, 
but typically gives raise to the dominant electronic correlations.
This is because inter-site interaction terms
yield mainly the Hartree contribution
if the number of neighbors is large \cite{MuellerHartmann89a}, and this
 is already included in LDA. The concept of Wannier functions
has hence found its use also in
materials specific many-body calculation with
the LDA+dynamical mean field theory (DMFT) 
approach \cite{Anisimov97a}.
The first LDA+DMFT calculations made use
of downfolded linearized muffin tin orbitals (LMTO)  \cite{LMTO1} 
and $N$th order muffin tin orbitals (NMTO) \cite{Andersen99}
which exploit the local character of the basis functions. 
For other bandstructure methods with
bases of extended orbitals a transformation
to Wannier functions is needed.

While the conversion from Bloch to Wannier functions is essentially
a Fourier transformation, there remains an arbitrariness in
choice of the phases of the Bloch functions; i.e.,
this conversion is not unique. This arbitrariness can be
used to construct localized orbitals.
Marzari and Vanderbilt \cite{Marzari97,Marzari01} developed a
procedure to fix the phases (and further 
${\mathbf k}$-dependent unitary transformations) so that
the resulting Wannier orbitals minimal spread around their
centers. Together with coworkers, they developed a corresponding 
program package wannier90 \cite{Wannier90}.

Since then the maximally localized Wannier functions (MLWF)
have been implemented for many DFT/LDA program packages
ranging from plane waves  to LMTO \cite{Lechermann06,Miyake08} 
and linearized augmented plane waves (LAPW)
in the FLAPW \cite{FLAPW} and FLEUR code \cite{FLEUR}.
Alternatives to MLWF have been proposed as well, including the recent 
Wannier function projection \cite{Ku02,Anisimov05} 
where the overlap of the
Bloch functions with the $d$- or $f$-orbitals within the atomic
or muffin tin spheres defines the Wannier functions.
This approach has been implemented for LMTO  \cite{Anisimov05} 
and  by Aichhorn {\em et al.}  \cite{Aichhorn09} for the FLAPW of the 
Wien2K code \cite{WIEN2k}. While this approach is somewhat
 easier to implement, the final Wannier functions are no longer
uniquely defined and may differ considerably. With a clever choice
of the functions to project onto,
results very similar to MLWF can be obtained.

In this paper, we present an interface \cite{w2w} between Wien2K ~\cite{WIEN2k} 
implementation of the FLAPW method and the 
wannier90 code \cite{Wannier90}. This iterface generates the
wannier90 input files (.eig, .mmn, .amn) from the Wien2K electronic structure.
Futhermore, we provide a code for the generation of MLWFs in  direct space, primarily useful for 
visualization.
The interface consists of FORTRAN90 programs  (tested for gfortran
and  ifort; hitherto not parallelized) with additional shell and python scripts to simplify
the work-flow.
A manual is available at  \cite{w2w}; the work-flow will be briefly
described in Section  \ref{computational details} 
The runtime of the main program~(w2w) was 
4.6 s for a simple test system such as SrVO$_3$ with 64 $\mb{k}$-points,
 running under Linux with a 64-bit Intel Xeon 3 GHz 4GB RAM.
For visualization, an output
to XCrysSDen \cite{XCrySDen} is also provided.

The following sections are organized as following:
In Section \ref{Theory}, we discuss the theoretical background
and equations, in particular the 
projection onto maximally localized Wannier function
(MLWF) in Section \ref{wan-intro} and the 
LAPW of Wien2K  in Section \ref{lapw}.
Some practical considerations are presented in  Section~\ref{practical considerations}.
The workflow of our Wien2Wannier interface is
outlined in Section~\ref{computational details}.
In   Section  \ref{results}, we present exemplarily some applications of the code.
We start with the simple bandstructure of SrVO$_3$ in Section \ref{results:SrVO3} to demonstrate
how the choice of the energy window affects the shape of WFs and the corresponding
tight-binding Hamiltonian. 
Sr$_2$IrO$_4$ with an 
idealized structure illustrates in Section \ref{results:Sr2IrO4} the construction of WFs in a system
with strong spin-orbit coupling leading to orbitals with complex phases.
In Section \ref{results:LaFeAsO}, LaFeAsO is used to show the utility of WFs for backfolding of 
the electronic bandstructures, in order to obtained the simplest possible ${\bf k}$-space
representation of the electronic structures.
Finally, in Section \ref{results:FeSb2},  we show  the
 application of the code to a material with low site-symmetry
 by the example of FeSb$_2$.

\section{\label{Theory}Theory} 
 
\subsection{\label{wan-intro} Wannier functions} 
Following Refs. \cite{Marzari97,Marzari01} we define the Wannier orbitals $w_{n\mb{R}}(\mb{r})$ in the 
${R}$'th unit cell through a unitary transformation $U^{(\mb{k})}$ of the Bloch eigenstates
$\psi_{n\mb{k}}(\mb{r})$ of the periodic Hamiltonian:
\begin{equation}
\label{eq:wan}
w_{m\mb{R}}(\mb{r})=
\frac{V}{(2\pi)^3}
\int_{BZ}e^{-i\mb{k}\cdot\mb{R}}
\left(\sum_nU^{(\mb{k})}_{nm}
\psi_{n\mb{k}}(\mb{r})\right).
\end{equation}
In the following the overall phases of the Bloch functions are assumed to be arbitrary, except
of the periodic gauge constraint $\psi_{m\mb{k}+\mb{G}}=\psi_{m\mb{k}}$ for $\mb{G}$ being
a reciprocal lattice vector. 
According to the Bloch theorem the function $\psi_{n\mb{k}}$ factorizes
into a product of the plane wave $\exp(i\mb{k}\cdot\mb{r})$ and a periodic function $u_{n\mb{k}}$:
\begin{equation}
\label{eq:psi}
\psi_{n\mb{k}}=e^{i\mb{k}\cdot\mb{r}}u_{n\mb{k}}.
\end{equation}
We are looking for that  $U^{(\mb{k})}$ which yields MLWFs,
i.e.,  minimizes the spread (variance) 
\begin{equation}
\sum_m \langle w_{m\mb{R}}(\mb{r})| \mb{r}^2|  w_{m\mb{R}}(\mb{r}) \rangle- \langle w_{m\mb{R}}(\mb{r})| \mb{r}|  w_{m\mb{R}}(\mb{r}) \rangle^2.
\end{equation}
 As the Wannier orbitals are expressed in terms of the Bloch 
wave functions in Eq.\ (\ref{eq:wan}), we need to
calculate the following overlap integrals
on a uniform $\mb{k}$-mesh in the Brillouin zone (BZ):
\begin{eqnarray}
\label{eq:mmn}
M_{mn}^{(\mb{k},\mb{b})}&=&\langle u_{m\mb{k}}|u_{n\mb{k}+\mb{b}} \rangle \label{eq:mmn1} \\
                        &=&\langle \psi_{m\mb{k}}|e^{-i\mb{b}\cdot\mb{r}}|\psi_{n\mb{k}+\mb{b}} \rangle \label{eq:mmn2}
\end{eqnarray}
For further details on how to determine the MLWF
from these  overlap integrals, we refer the reader to \cite{Marzari97,Marzari01}.

Moreover, for the disentanglement procedure described in
\cite{Marzari01} and  a good starting guess, also overlap integrals with a set of fixed trial orbitals $g_n$ 
\begin{equation}
A_{mn}^{(\mb{k})}=\langle \psi_{m\mb{k}}|g_n\rangle
\end{equation}
have to be calculated. The evaluation of $M_{mn}^{(\mb{k},\mb{b})}$ and $A_{mn}^{(\mb{k})}$ in the
LAPW basis is the subject of the rest of this section.

\subsection{\label{lapw} Linearized augmented plane-waves}
In the LAPW method, the direct space is divided into two regions: 
the interstitial space I and the non-overlapping muffin-tin (MT) spheres 
S$_{\beta}$ around the nuclei at ${\bf R}_{\beta}$. 
The Bloch functions are expanded into plane waves in I and into partial atomic waves in the MT spheres 
\begin{equation}
\label{eq:lapw}
\psi^{\sigma}_{\mb{k}}(\mb{r})=
\left\{
\begin{array}{l@{\quad}l}
\frac{1}{\sqrt{V}}\sum_{\mb{G}}C^{\sigma}_{\mb{k}}(\mb{G})e^{{\rm i}\left({\bf k}+{\bf G}\right){\bf r}} &
{\bf r}\in {\rm I} \\
\sum_{\ell m}\left(a_{\mb{k},\ell m}^{\beta\sigma}
u_{1,\ell}^{\beta\sigma}(r_{\beta}) +
b_{\mb{k},\ell m}^{\beta\sigma}
\dot{u}_{1,\ell}^{\beta\sigma}(r_{\beta}) +
c_{\mb{k},\ell m}^{\beta\sigma}
{u}_{2,\ell}^{\beta\sigma}(r_{\beta})
\right)
Y_{\ell m}(\hat{\bf r}_{\beta}) &  {\bf r}\in{\rm S}_{\beta}, 
\end{array}
\right.
\end{equation}
where $Y_{\ell m}$ are the spherical harmonics and $u_1$, $\dot{u}_1$, and $u_2$ are numerically determined functions
of the radial variable $r_{\beta}=|{\bf r}-{\bf R}_{\beta}|$, which are defined separately
for each MT sphere $\beta$, orbital quantum number $\ell$ and, in spin-polarized calculations,
spin projection $\sigma$. For simplicity we do not show the dependence of the 
expansion coefficients $C^{\sigma}_{\mb{k}}$, $a_{\mb{k},\ell m}^{\beta\sigma}$, $b_{\mb{k},\ell m}^{\beta\sigma}$, and
$c_{\mb{k},\ell m}^{\beta\sigma}$ on the band index. 
Both LAPWs and augmented plane waves plus local orbitals (APW+lo) 
basis sets can be used for our interface.
While further details can be found
in Ref. \cite{Singh94} we point out two features of the expansion:
(i) it is not possible to obtain the periodic part $u_{n\mb{k}}$ inside the spheres  in a simple straight-forward
way from the above expansion of the Bloch function and therefore we used Eq.\ (\ref{eq:mmn2}) in actual
calculations; (ii) the Bloch function is uniquely defined by $C^{\sigma}_{\mb{k}}(\mb{G})$ 
since the other expansion coefficients are in fact  functions of $C^{\sigma}_{\mb{k}}(\mb{G})$,
defined by the continuity conditions at the surfaces of the MT spheres.

{\it The calculation of $M_{mn}^{(\mb{k},\mb{b})}$} naturally splits into the interstitial part
and the MT part. To evaluate (\ref{eq:mmn2}) inside the MT sphere $\beta$, the plane-wave
$\exp(-i\mb{b}\cdot\mb{r})$ is expanded into products of spherical harmonics and Bessel
functions using the well known formula
\begin{equation}
\label{eq:bess}
e^{-i\mb{b}\cdot\mb{r}}=4\pi e^{-i\mb{b}\cdot\mb{R}_{\beta}}\sum_\ell i^{\ell}j_{\ell}(br_{\beta})\sum_m \bar{Y}_{\ell m}(-\hat{\mb{b}})
Y_{\ell m}(\hat{\mb{r}}_{\beta}).
\end{equation}

Substituting Eqs.\ (\ref{eq:lapw}) and (\ref{eq:bess}) into Eq.\ (\ref{eq:mmn2}), one obtains an 
expression 
for the contribution of $M_{mn}^{(\mb{k},\mb{b})}$ within the
MT spheres in terms of products of radial integrals   
$\int_0^{R_{\beta}}dr\; r^2 u_{p,\ell_1}^{\beta\sigma}(r)j_{\ell_2}(br)u_{q,\ell_3}^{\beta\sigma}(r)$;
angular integrals $\langle\langle \bar{Y}_{\ell_1 m1}Y_{\ell_2 m2}Y_{\ell_3 m_3}\rangle\rangle$, known as Gaunt numbers; and an additional
weight factor given by
 the corresponding products of the expansion coefficients.

The interstitial contribution $M_{mn}^{(\mb{k},\mb{b})}$ on the other hand 
reads
\begin{equation}
\langle \frac{1}{V}\sum_{\mb{G}\mb{G}'} \bar{C}^m_{\mb{k}}(\mb{G}')C^n_{\mb{k}+\mb{b}}(\mb{G})\langle e^{i(\mb{G}-\mb{G}')\cdot\mb{r}}\rangle_I,
\end{equation}
where the $\langle\rangle_I$ integration runs over the unit cell with the MT spheres removed, which gives
\begin{equation}
\begin{array}{c}
\langle e^{i(\mb{G}-\mb{G}')\cdot\mb{r}}\rangle_I=V\delta_{\mb{G},\mb{G'}}-3V_{\beta}\sum_{\beta}e^{-i(\mb{G}-\mb{G}')\cdot\mb{R}_{\beta}}
\frac{\sin(x)-x\cos(x)}{x^3} \\
\mbox{where}\quad x=\rho_{\beta}|\mb{G}-\mb{G}'|, \quad V_{\beta}=\frac{4}{3}\pi \rho_{\beta}^3,
\end{array}
\end{equation}
and $\rho_{\beta}$ is the radius of the MT sphere at $\mb{R}_{\beta}$.

The periodic gauge has to be enforced when the integration in Eq.\ (\ref{eq:wan}) wraps around the Brillouin zone boundary,
namely a reciprocal vector $\mb{G}$ must be chosen such that $\mb{k}+\mb{b}-\mb{G}$ lies within the Brillouin zone.

{\it The construction of $A_{mn}^{(\mb{k})}$} is rather simple if we allow trial orbitals to be non-zero only
inside the MT sphere, i.e. of the form
\begin{equation}
\label{eq:target}
g_n=\sum_{\beta} \sum_{\ell m} g^{\beta}_{\ell m}(n)u^{\beta}_{1,\ell}(r_{\beta})Y_{\ell m}(\hat{\mb{r}}_{\beta}),
\end{equation}
where the list of non-zero coefficients $g^{\beta}_{\ell m}(n)$ is provided as input. 
Using (\ref{eq:lapw}) we obtain
\begin{equation}
A_{n'n}^{(\mb{k})}=\sum_{\beta} \sum_{\ell m} \left( \bar{a}^{n'}_{\mb{k}, \ell m} \langle u^{\beta}_{1,\ell}|u^{\beta}_{1,\ell} \rangle_r +
\bar{b}^{n'}_{\mb{k}, \ell m} \langle \dot{u}^{\beta}_{1,\ell}|u^{\beta}_{1,\ell} \rangle_r +
\bar{c}^{n'}_{\mb{k},\ell m} \langle u^{\beta}_{2,\ell}|u^{\beta}_{1,\ell} \rangle_r \right) g^{\beta}_{\ell m}(n),
\end{equation}
where $\langle\rangle_r$ denotes the radial integral with $r^2dr$.

\subsection{\label{practical considerations}Practical considerations}
In the following we briefly describe the steps taken in computing WFs. We start by generating a uniform {\bf k}-mesh
in the Brillouin zone (running KGEN with only the identity matrix in the list of symmetry operations) followed
by generating the eigenstates using LAPW1. In case of calculations with spin-orbit coupling
LAPWSO should be executed. Spin-orbit calculation must be run as formally spin-polarized even if the polarization
is zero. Before running the wien2wannier interface code the initialization run of wannier90 must be executed
to generate the case.nnkp file. The inputs to wien2wannier specifies the bands in the initial
Hilbert space, the number of the bands in the target Hilbert space and the expansions (\ref{eq:target}) of the target
orbitals. The wien2wannier generates the necessary files case.eig, case.mmn and case.amn, which
serve as an input for the MLWF construction by wannier90. In case of spin-orbit coupled calculations,
a separate run of wien2wannier is executed for each of the two components of the spinor wave function.
The corresponding elements of the resulting case.mmn(amn)  files are added up to form
the total overlap matrices between the spinor functions. The postprocessing of the
wannier90 results is, to a large extent, independent of the  bandstructure code. However,  for plotting
the WFs in  direct space information on
 underlaying basis functions is necessary. We have modified the LAPW7 code of the Wien2K package to generate
direct space WF maps, which uses the transformation between the initial Bloch states
and resulting WFs extracted from the wannier90 calculation.     

\subsection{\label{computational details}Computational details}
The calculations reported in the following Section were performed with Wien2k \cite{WIEN2k} code employing
the LDA exchange-correlation functional  
for SrVO$_3$, Sr$_2$IrO$_4$, LaFeAsO and the generalized gradient correction (GGA)  for FeSb$_2$; in all cases  
the APW+lo basis set \cite{SjostedtSSC00} was employed.
To construct the Wannier orbitals we followed the sequence:
(i) self-consistent bandstructure calculation using the irreducible part of the BZ;
(ii) generation of a uniform k-mesh throughout the entire BZ (wien2k) and of the corresponding
$\mb{k},\mb{k}+\mb{b}$-list (wannier90);
(iii) calculation of the Bloch eigenstates and eigenvalues on the new k-mesh (wien2k);
(iv) evaluation of the $M_{mn}^{(\mb{k},\mb{b})}$ and $A_{mn}^{(\mb{k})}$ elements (wien2wannier);
(v) generation of MLWFs (wannier90);
(vi) post-processing of MLWFs to generate tight-binding bandstructures, hopping integrals, plots of
Wannier orbitals (wannier90+wien2k).
In case of calculations with spin-orbit coupling the Bloch functions have two non-zero
components, indexed with the spin quantum number, with a definite mutual phase. Once the
two components are generated, the contributions to $M_{mn}^{(\mb{k},\mb{b})}$ and $A_{mn}^{(\mb{k})}$
of each component are computed separately along the lines of the previous section and added up at
the end.
To plot the Wannier orbitals we extract the $U^{(\mb{k})}_{mn}$ matrices of Eq. (\ref{eq:wan}) from 
the wannier90 processing and combine it with the wien2k utility for the generation of the 
Bloch state on direct space grids.

\section{\label{results}Results}

\subsection{\label{results:SrVO3}SrVO$_3$}
\begin{figure}
\begin{center}
\includegraphics[height=0.6\columnwidth,clip]{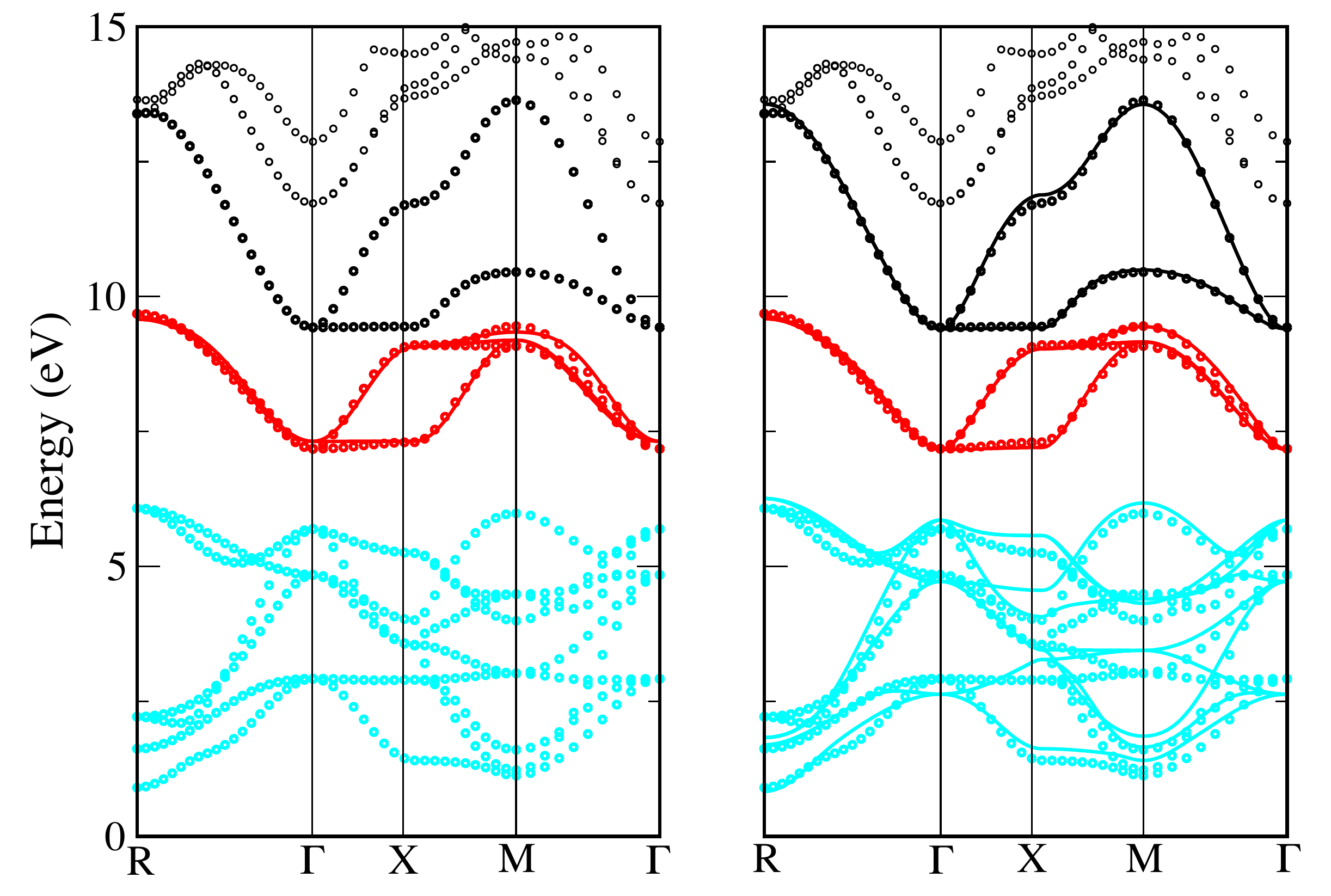}
\caption{\label{fig:srvo3-band}(color online) LDA band structure of SrVO$_3$ (circles) 
with dominant contributions marked with color: O-$p$ (blue), V-$d$-$t_{2g}$ (red), and
V-$d$-$e_{g}$ (black).
Tight-binding bands obtained from MLWFs for (i)[left panel] only the $t_{2g}$ band  (ii)[right panel]  all  V-$d$ and O-$p$ bands,
  with spatial cut-offs  as described in the text (the Fermi level is at 8.16~eV).}
\end{center}
\end{figure}

\begin{figure}
\begin{center}
\includegraphics[height=0.7\columnwidth,angle=270,clip]{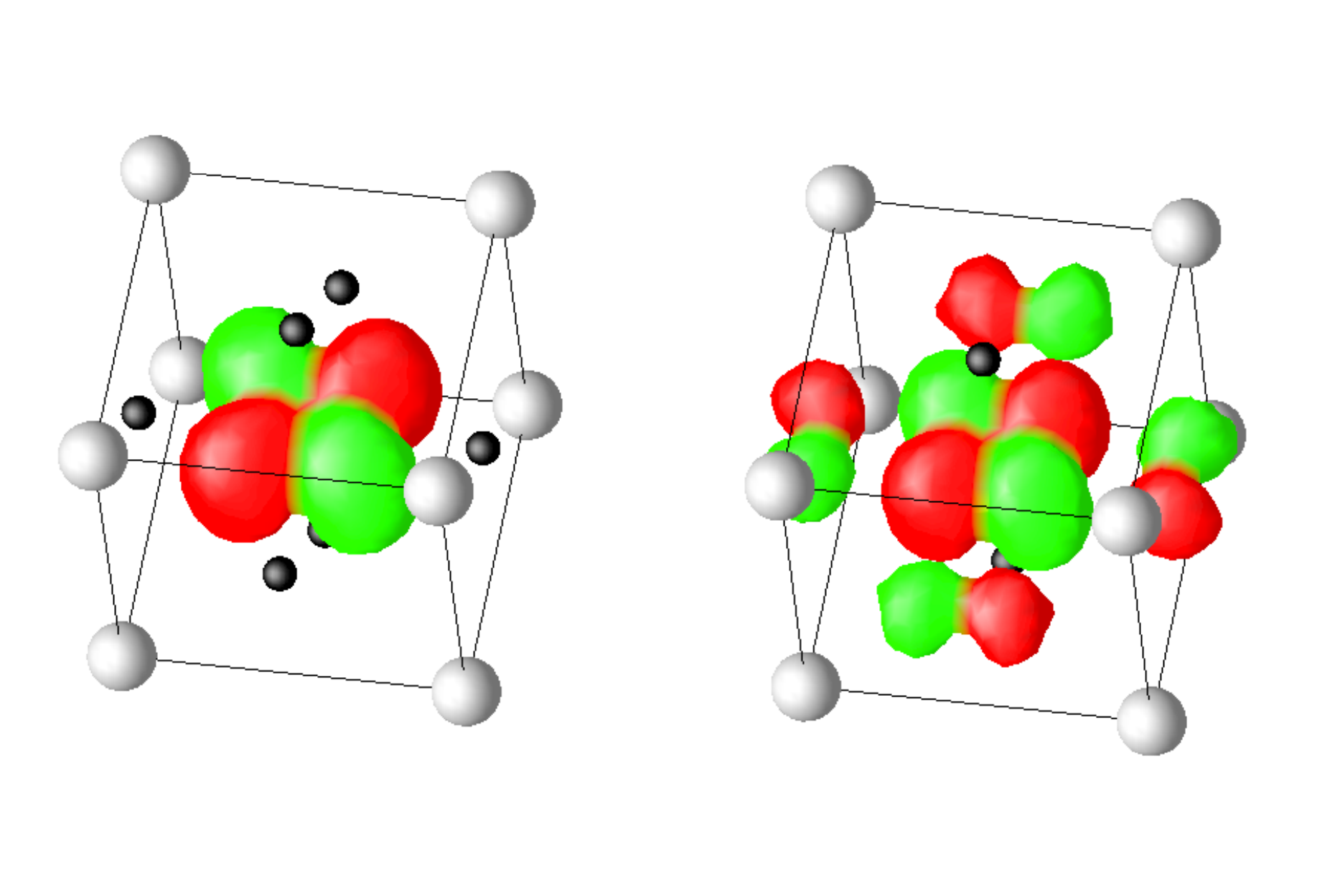}
\caption{\label{fig:srvo3-wan} (color online) 
LDA band structure of Sr$_2$IrO$_4$ (back circles) together
with the MLWF tight-binding fit to the $J=1/2$ bands for different direct space cut-offs; the Fermi level
is at 8.2~eV. (left)
The $xy$ Wannier orbital plotted as an isosurface
of the charge density $|w(\mb{r})|^2$and colored by the sign of $w(\mb{r})2$. The left panel
corresponds to the large energy window (ii) and the right panel to the small energy window (i) for
the same isovalue. (right)}
\end{center}
\end{figure}

SrVO$_3$ is a  thoroughly studied material with a rather simple band structure
(see Fig. \ref{fig:srvo3-band})
consisting of isolated groups of bands derived from O-$p$, V-$d$-$t_{2g}$ and V-$d$-$e_g$ orbitals. Therefore it is very well suited as a testing case.
We focus on the V-$d$-$t_{2g}$ states, which (being partially filled) are of  most interest.
We compare two different settings, i.e., consisting of two choices of the Hilbert space to be represented by the Wannier functions.
Namely, the space spanned by (i) $t_{2g}$ bands only
 and (ii) all the V-$d$ and O-$p$ bands.
In both cases, the $t_{2g}$ Wannier functions are orthogonal to the O-$p$ Bloch or Wannier
states. In the former case (i) also the hopping integral between the $t_{2g}$ and O states is still zero
at the price of more extended Wannier orbitals with a substantial weight at the O sites.
Naturally, the larger energy window (ii) clearly allows the construction of more localized WFs
as demonstrated in Fig. \ref{fig:srvo3-wan}. The Wannier orbitals (i) have visible density
on the neighboring O sites, reflecting the mixed character of the $t_{2g}$ bands.
In contrast, the orbitals (ii) constructed from the larger energy window do not have any appreciable (visible) 
density  at O sites (on the scale of the figure) since the latter would be  now assigned to the O-$p$
orbitals, which are explicitly presented by O-centered Wannier orbitals.

The spatial extent of the Wannier orbitals is also reflected in the hopping
integrals, which were calculated by the wannier90 code.  With the small energy window (i) the $t_{2g}$ bands are well described when at least nn- and nnn-hoppings
is considered (see Fig. \ref{fig:srvo3-band}), which in the $\{xy,yz,zx\}$ basis read
\begin{equation}
t_{100}[\mbox{meV}]=\left(\begin{array}{ccc} -268 & 0 & 0 \\ 0 & -30 & 0 \\ 0 & 0 & -268 \end{array} \right),
\quad\quad t_{101}[\mbox{meV}]=\left(\begin{array}{ccc}  7 & 10 & 0 \\ 10 & 7 & 0 \\ 0 & 0 & -93 \end{array} \right).
\end{equation}
 The remaining directions follow from symmetry considerations. The
longest nnn  $t_{2g}$-$t_{2g}$ hopping
 corresponds to a length of  of 5.4~\AA.
Using the more localized orbitals (ii) we can achieve similar accuracy (see Figure \ref{fig:srvo3-band}) by considering
only V-V nn-hopping and V-O nnn-hopping, which translate into a direct spatial cut-off of only 4.3~\AA. The obvious
price to be paid are  larger matrices ($14\times14$ for (ii) vs $3\times3$ in case (i)).
We point out when no spatial cut-offs are introduced, both choices (i) and (ii) represent the $t_{2g}$ bands
to the same arbitrarily high accuracy, determined by the size of the k-mesh used to construct the
MLWFs.

\begin{figure}
\centering
\includegraphics[height=0.5\columnwidth,clip]{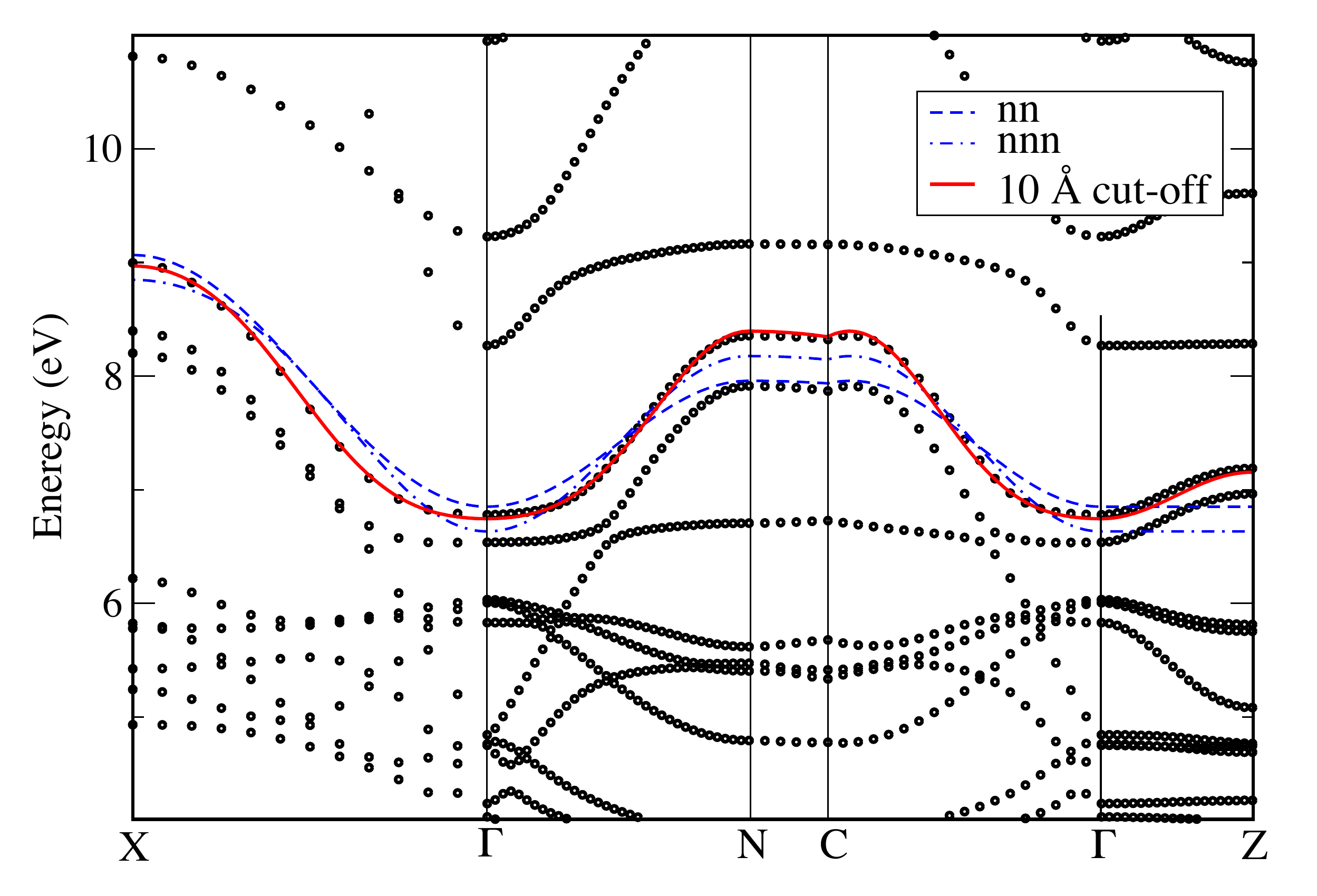}
\caption{\label{fig:sr2iro4-band}(color online) LDA band structure of Sr$_2$IrO$_4$ (back circles) together with the MLWF tight-binding fit to the $J=1/2$ bands for different direct space cut-offs (the Fermi level is at 8.2~eV).}
\end{figure}
\subsection{\label{results:Sr2IrO4}Sr$_2$IrO$_4$}
\begin{figure}
\centering
\includegraphics[height=0.4\columnwidth,angle=270,clip]{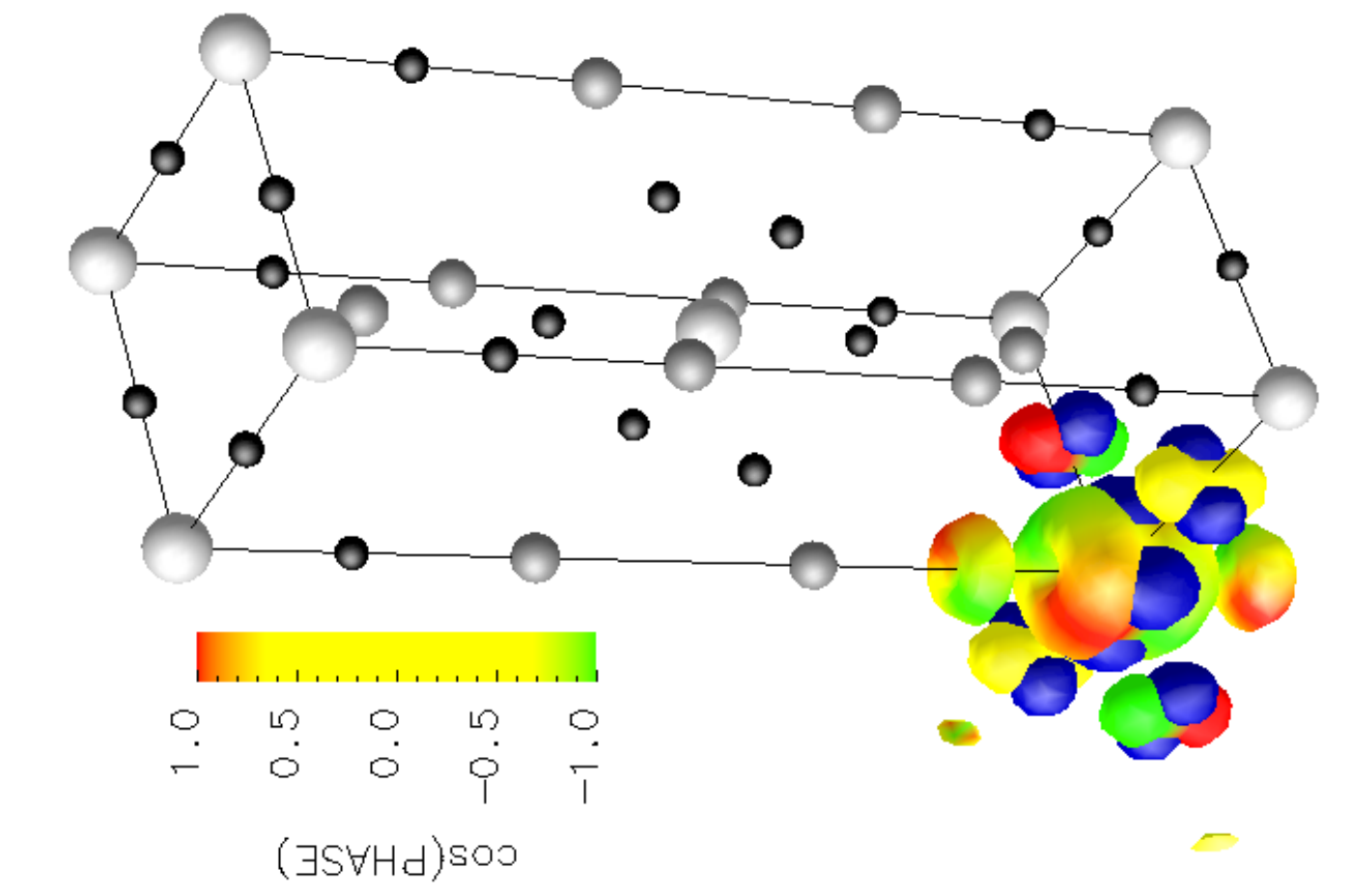}
\caption{\label{fig:sr2iro4-wan}(color online)  Ir $J=1/2$ Wannier orbital $w_{+,[0,0,0]}(\mb{r})$ visualized
as a $|w|^2$ isosurface. The almost real $\downarrow$-spin component is blue, the $\uparrow$-spin
component is colored with cosine of its phase (red=real positive, green=real negative, yellow=imaginary), see color legend bar.}
\end{figure}

Sr$_2$IrO$_4$ has been recently the subject of intense investigations  due to the close
connection between the spin-orbit coupling and its Mott-insulating ground state.
We use it  as a example of a material where spin-orbit coupling substantially modifies
the band structure and leads to Wannier orbitals, in which both spin projections are mixed.
For sake of simplicity we have performed the calculations using an idealized double-perovskite
structure, while the real material is characterized by a  tilting of the IrO$_6$ octahedra.
In the following, we discuss the LDA band structure of Sr$_2$IrO$_4$, which
represents a crude approximation to correlation effects
due to local interactions. Within this picture the electronic structure (see Fig. \ref{fig:sr2iro4-wan})
can be understood by considering the crystal-field splitting, spin-orbit coupling and the 
inter-site hopping via the manifold of Ir-$d$ bands.
The crystal-field splitting, being the largest of the three, opens a gap between the $t_{2g}$ and $e_g$ bands, rendering the 
latter empty, while the formed accommodate one hole per Ir atom. The $t_{2g}$ orbitals may be labeled
with a pseudo-spin $I=1$. The spin-orbit coupling further splits the $t_{2g}$ manifold into a
quadruplet and doublet with pseudo-spin $J=3/2$ and $J=1/2$, respectively. Since the
spin-orbit splitting is rather large the inter-site hopping leads only to moderate mixing
of the states with different $J$. Therefore we may expect the isolated band doublet at the top of the
$t_{2g}$ manifold to be predominantly of $J=1/2$ character. We hence construct the MLWFs for these two bands.
\begin{figure}
\begin{center}
\includegraphics[height=0.5\columnwidth,clip]{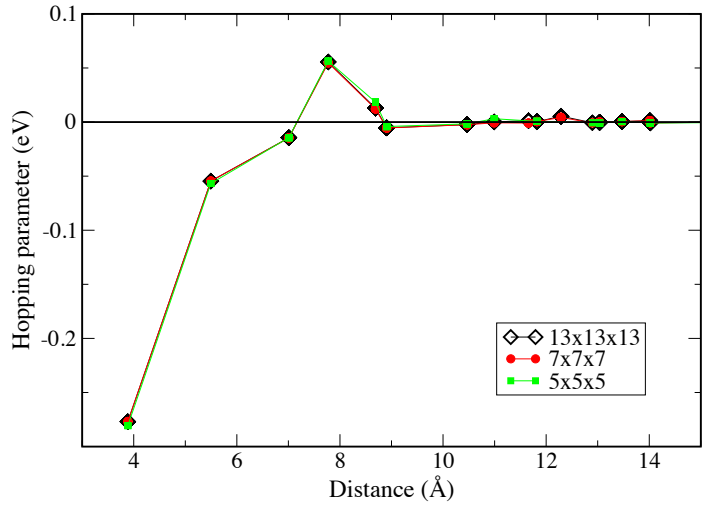}
\caption{\label{fig:sr2iro4-hop}(color online) The tight-binding hopping integrals
of the one-band model  as a function of inter-site distance for different BZ grid sizes
(the lines are guides for eyes).}
\end{center}
\end{figure}

For testing purposes, we have prepared Bloch states on k-grids of sizes $3\times3\times3$, $5\times5\times5$, and
$7\times7\times7$.  As trial functions we have used the $J_{eff}=1/2$ orbitals, which adopt the following
spinor form with respect to the local coordinate axes pointing towards  the $O$ atoms:
\begin{equation}
|\phi_+\rangle\sim\left(\begin{array}{c} -2Y_{21} \\ Y_{2-2}-Y_{22} \end{array}\right),\quad |\phi_-\rangle\sim\left(\begin{array}{c} Y_{2-2}-Y_{22} \\ 2Y_{2-1} \end{array}\right),
\end{equation}
where we omit the unimportant normalization factor.
The results for the three different k-grids were almost identical with minor deviations only for the 
smallest grid. Due to a clever choice of trial functions, the original spread of the Wannier orbitals as defined in wannier90 actually changed by 
less than $0.5\%$ during the MLWF optimization. This is consistent with the
fact that the resulting MLWFs are essentially $J_{eff}=1/2$ functions as  shown in Fig. \ref{fig:sr2iro4-wan}.
In particular, the Wannier orbital, to a very good approximation, consists of a real $xy$ orbital in one spin channel and a
complex $(x\pm iy)z$ orbital in the other ($\pm$) spin channel. Note that the relative phase of the two
components is not arbitrary and the corresponding charge density (sum of the spin components)
has (approximate) cubic symmetry as expected for a $J_{eff}=1/2$ orbital.

In Fig. \ref{fig:sr2iro4-wan} we show convergence of the tight-binding bands to the original band structure
with increasing maximum hopping distance. The band dispersion is governed
by the nearest- and next-nearest-neighbor hopping and the bands are essentially converged when
five coordination spheres are considered, which includes also the out-of-plane hoppings
giving rise to the $z$-axis dispersion. The hopping amplitudes are summarized in Fig. \ref{fig:sr2iro4-hop}.

\subsection{\label{results:LaFeAsO}LaFeAsO}

\begin{figure}
  \begin{center}
    \includegraphics[height=0.5\columnwidth,clip]{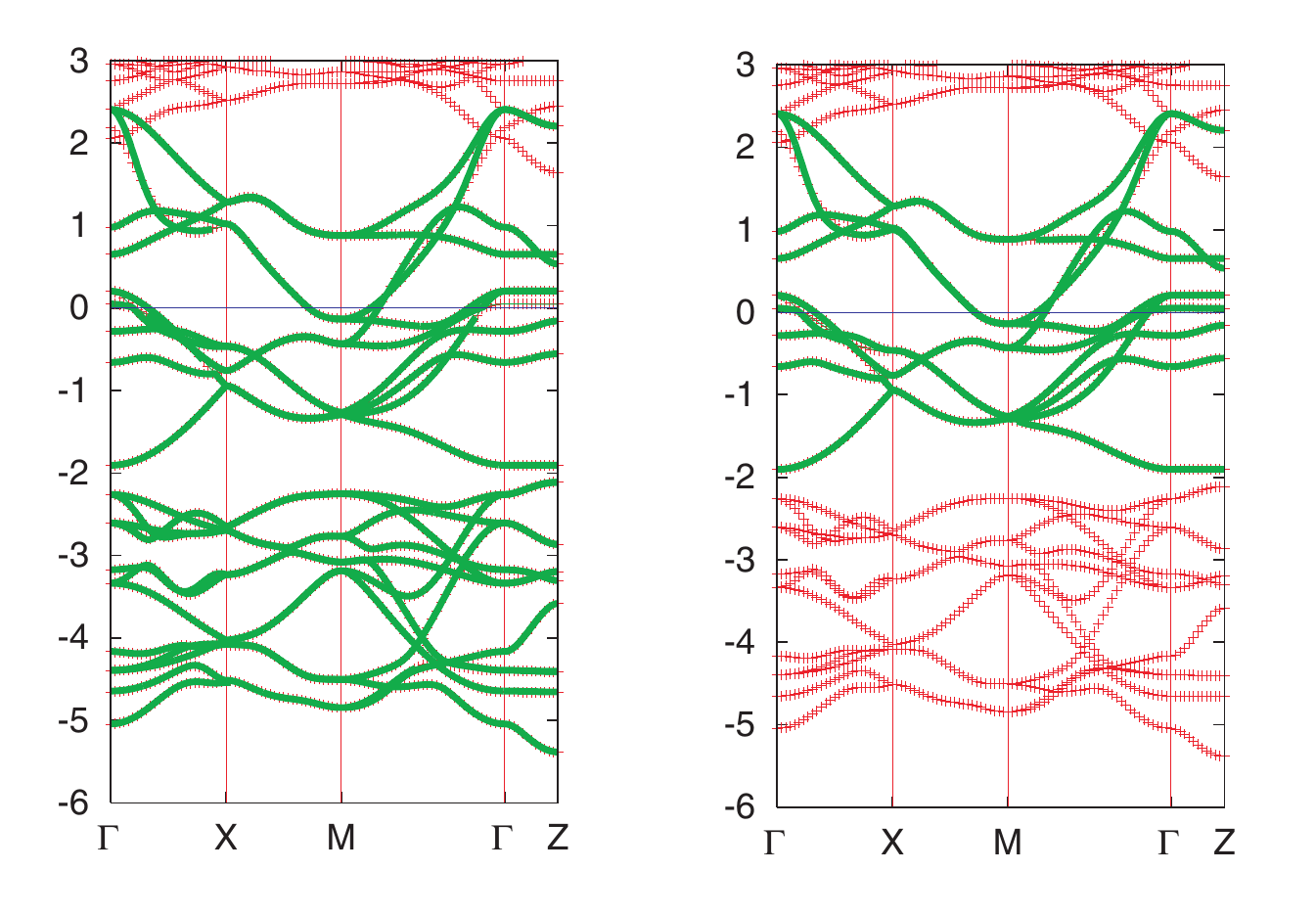}
  \end{center}
\caption{\label{fig:LaFeAsO-band}
(color online) LDA band structure of LaFeAsO (red cross symbols) 
compared with tight-binding bands obtained from MLWFs (green solid lines)
for the $dpp$ model (left) and the $d$ model (right); the Fermi level is set to zero.
}
\end{figure}

In 2008, Kamihara {\it et al.} discovered superconductivity in F-doped LaFeAsO 
\cite{Hosono}, immediately  followed by the same finding in related materials. This opened a new field of research inviting many theoretical
approaches to be applied, which in turn called for simplified models of the electronic structure capturing
the essential chemistry of these systems.
These new superconductors commonly have two-dimensional
iron pnictide or iron chalcogenite layers, and 
the low-energy electronic structure around the Fermi level
($E_F$) consists of heavily entangled bands dominated by the Fe 3$d$ states
(see Fig.\ref{fig:LaFeAsO-band}).
While the multi-band character renders tight-binding fit impractical, the
exact transformation using Wannier functions provides a controlled
way of constructing tight-binding models representing the bandstructure
to arbitrary accuracy, see Fig.\ref{fig:LaFeAsO-band} and
 \cite{KurokiPRL08,CaoPRB08,Nakamura08,Vildosola08,Anisimov09,Miyake10,Arita09,Ikeda10}.

\begin{figure}
  \begin{center}
    \includegraphics[height=0.5\columnwidth,clip]{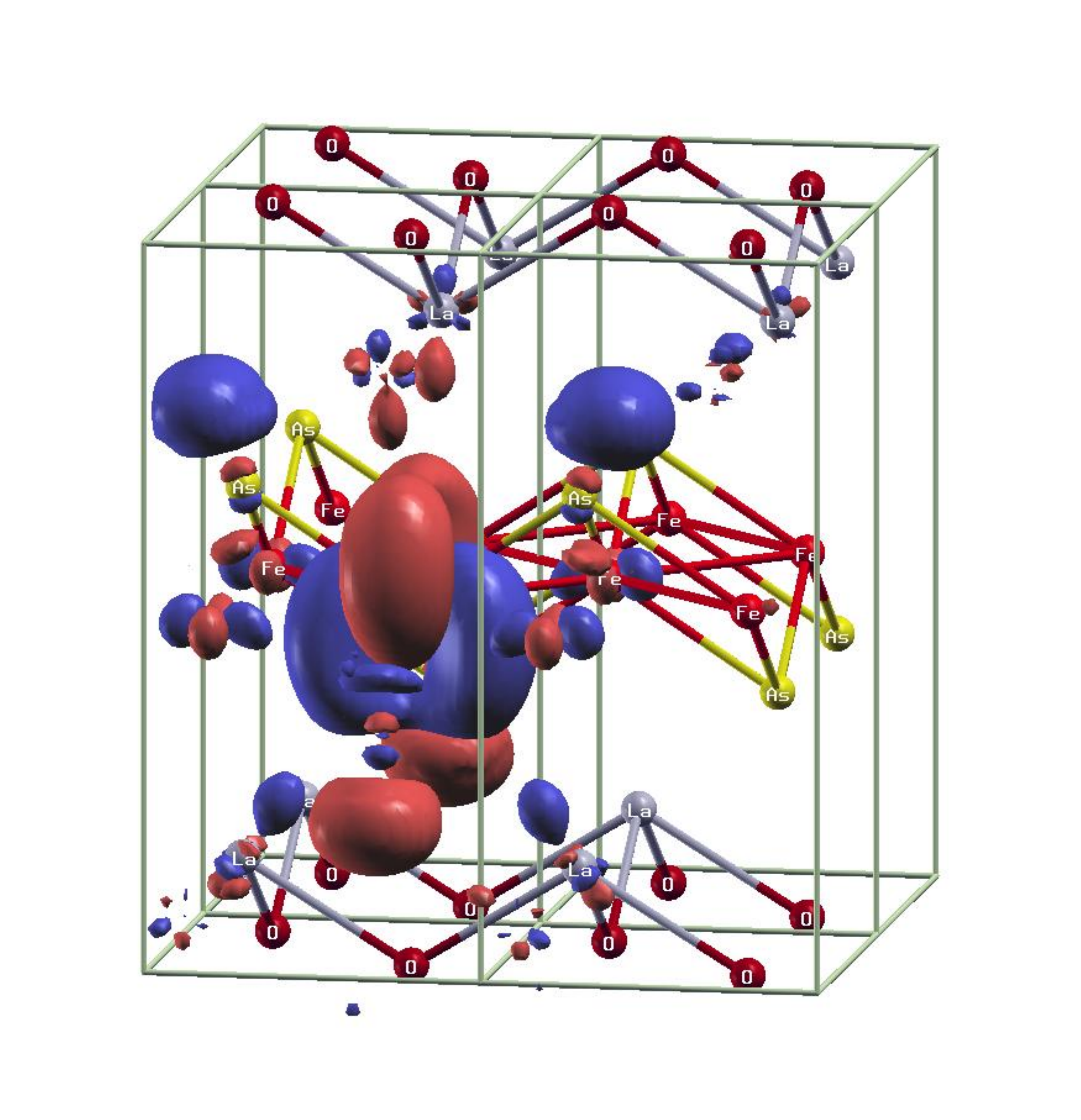}
  \includegraphics[height=0.5\columnwidth,clip]{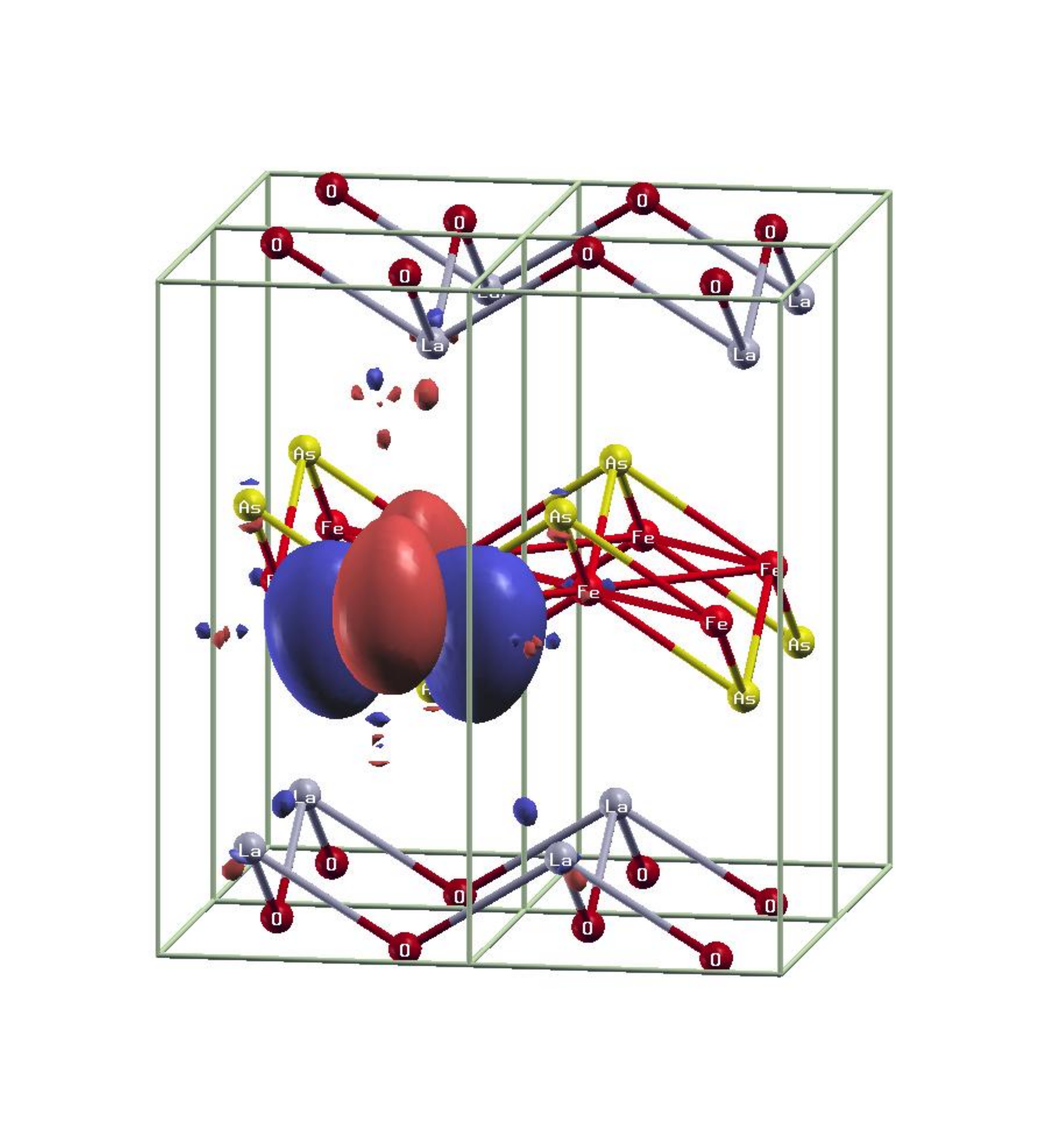}
  \end{center}
\caption{\label{fig:LaFeAsO-Wan}
(color online) Isosurface contours of the $x^2$-$y^2$ MLWO 
for the $dpp$ model (left) and the $d$-only model (right).
The amplitudes of the contour surface are +0.07 \AA$^{-3/2}$ (red)
and -0.07 \AA$^{-3/2}$ (blue).
}
\end{figure}

Two types of tight-binding models have been considered for iron pnictides:
the $d$ model  on the basis of  Fe 3$d$-like orbitals,
and $dp$ or $dpp$ model using the Fe 3$d$, pnictogen/chalcogen $p$,
and oxygen $p$ orbitals. 
In both cases the corresponding bands are reproduced exactly,
as shown in Fig.\ref{fig:LaFeAsO-band}, but the behavior of the two models
differs if electron-electron interactions are explicitly included.
As in the case of SrVO$_3$, the shape and spread of the Fe $3d$ orbitals
depends on whether the $d$ or $dp$ model is used, see Fig.\ref{fig:LaFeAsO-Wan}
and Table \ref{spreads}.



\begin{figure}
  \begin{center}
    \includegraphics[height=0.5\columnwidth,clip]{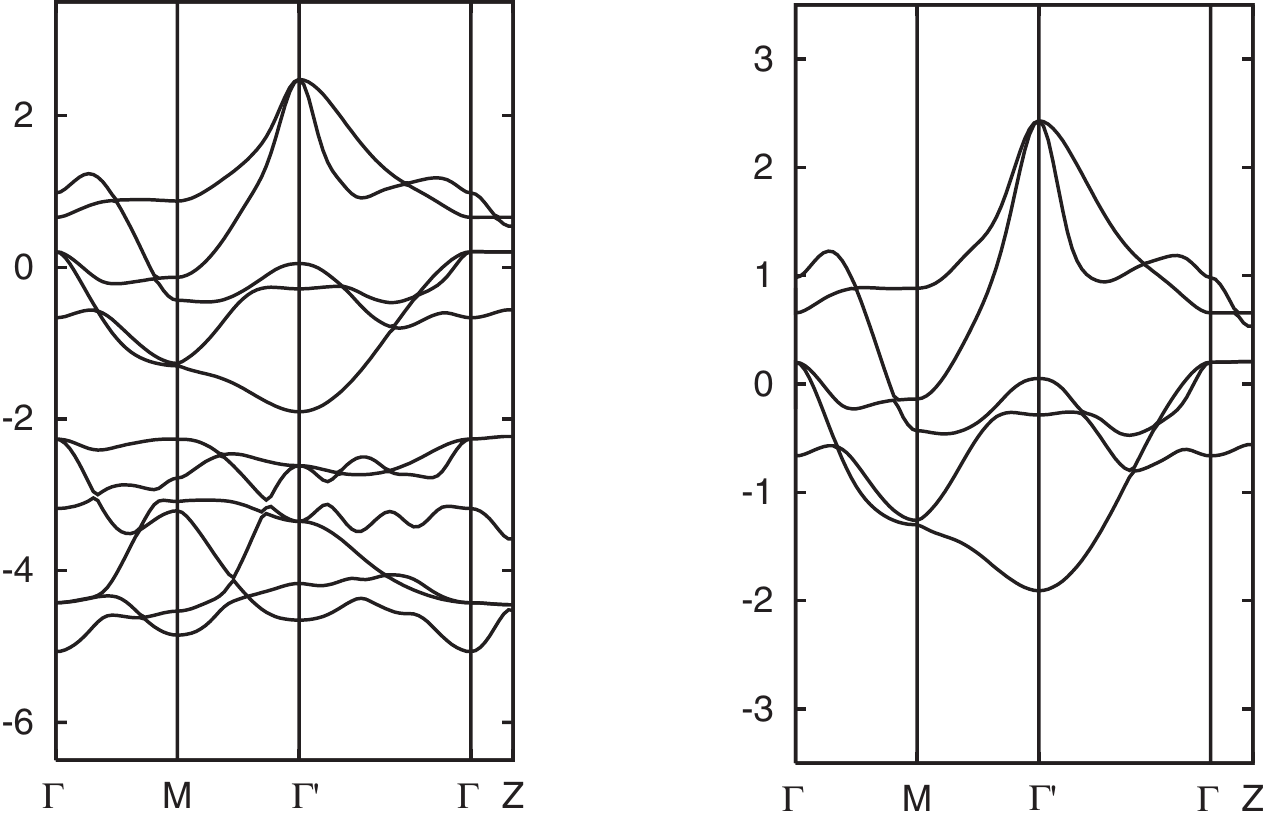}
  \end{center}
\caption{\label{fig:unfold}
Band dispersion of the $dpp$ (left) and $d$ model (right) for LaFeAsO with 
the first Brillouin zone being extended to the second Brillouin
zone, so that the unit cell contains only one Fe atom. 
Both $\Gamma$ and $\Gamma'$ points correspond to $\Gamma$ in the
original band structure in Fig.\ref{fig:LaFeAsO-band}.
}
\end{figure}

One of the advantages of  a tight-binding model, which represents the band
dispersion of {\it ab initio} calculation, is that we can sometimes ``unfold'' the
Brillouin zone.
The unit cell of LaFeAsO contains two Fe and As
atoms, where one of the As atoms sits above and the other below the
Fe plane. Therefore, there are ten Fe 3$d$ bands. It turns out that the translation symmetry
of the tight-binding model is higher and the bandstructure can be unfolded to a
five band scheme, corresponding to an effective unit cell with one Fe site.
The corresponding bandstructure is shown in Fig.\ref{fig:unfold}.
The missing translation symmetry between the two Fe sites of the crystallographic unit cell means that the WFs on the two sites are not connected by a simple
translation, but by a more general symmetry operation. Obviously, not all 
operators 
have the property
 of the tight-binding Hamiltonian that this 
does not matter; and therefore the five band model 
cannot replace the ten band one in general.

\begin{table}
\begin{center}
\begin{tabular}{|r|r|r|r|r|r|r|r|r|r|r|r|}
   & $z^2$ & $xz$ & $yz$ & $x^2$-$y^2$ & $xy$ & As-$pz$ & As-$px$ & As-$py$ & O-$pz$
& O-$px$ & O-$py$ \\
\hline
dpp& 1.82  & 2.13 & 2.13 & 2.41 &        1.74 &         &         &         &       
&        &        \\
dpp& 1.08  & 1.34 & 1.34 & 1.30 &        1.01 & 1.93    & 1.98    & 1.98    & 1.27  
& 1.30   & 1.30   \\
\end{tabular}
\end{center}
\caption{Spread of maximally localized Wannier functions for $d$ model and $dpp$
model (in units of \AA).
}
\label{spreads}
\end{table}


In table \ref{transfer}, we list the hopping parameters of the
$d$ model. They are very similar to previous studies
such as Ref. \cite{KurokiPRL08}, Ref. \cite{Miyake10},
of Ref.\cite{Ikeda10}. It should be noted that
since Fe 3$d$ bands hybridize with La 5$d$, there can be
slight difference, depending on whether 
we use the so-called ``inner'' window or not,  see \cite{Ikeda10}.

\begin{table}
\begin{center}
\begin{tabular}{c|rrrrrccc}
\small($\mu$,$\nu$)}{\small $[\Delta x, \Delta y]$
 &[1,0]&[1,1]&[2,0]&[2,1]&[2,2]&$\sigma_y$& $I$& $\sigma_d$\\ 
\hline
  ($z^2$,$z^2$) &   $-$66&   $-$8&  $-$33&     17 &  $-$15 &   +    & +   &+  \\
  ($z^2$,$xz$) &   $-$72&      0&      0&   $-$2 &      0 &$-$($z^2$,$yz$)& $-$ &$-$\\
  ($z^2$,$yz$) &      72& $-$144&      0&   $-$3 &  $-$27 &$-$($z^2$,$xz$)& $-$ &+  \\
  ($z^2$,$x^2$-$y^2$) &       0&    160&      0&      9 &  $-$15 &   $-$  & +   &+  \\
  ($z^2$,$xy$) &  $-$297&      0&   $-$3&  $-$20 &      0 &    +   & +   &$-$\\
  ($xz$,$xz$) &  $-$198&    133&      5&   $-$6 &      2 &+($yz$,$yz$)  & +   &+  \\
  ($xz$,$yz$) &     133&      0&     24&  $-$16 &      0 &    +   & +   &$-$\\
  ($xz$,$x^2$-$y^2$) &     167&      0&      0&     14 &      0 &+($yz$,$x^2$-$y^2$)
 & $-$ &$-$\\
  ($xz$,$xy$) &  $-$252&    137&      0&   $-$9 &      8 &$-$($yz$,$xy$)& $-$ &+  \\
  ($yz$,$yz$) &  $-$198&    321&      5&  $-$24 &     67 &+($xz$,$xz$)  & +   &+  \\
  ($yz$,$x^2$-$y^2$) &     167&     20&      0&     17 &      4 &+($xz$,$x^2$-$y^2$)
 & $-$ &+  \\
  ($yz$,$xy$) &     252&      0&      0&     26 &      0 &$-$($xz$,$xy$)& $-$ &$-$\\
  ($x^2$-$y^2$,$x^2$-$y^2$) &     154&    118&  $-$25&  $-$30 &  $-$26 &   +    & + 
 &+  \\
  ($x^2$-$y^2$,$xy$) &       0&      0&      0&  $-$11 &      0 &  $-$   & +   &$-$\\
  ($xy$,$xy$) &     313&  $-$68&  $-$18&      2 &      1 &   +    & +   &+  \\
\end{tabular}
\end{center}
\caption{Hopping integrals $t(\Delta x, \Delta y; \mu, \nu)$ 
in units of meV. $[\Delta x, \Delta y]$ denotes the in-plain 
hopping vector being different for different columns 
and $(\mu,\nu)$ the orbitals being different for the rows. The last
three columns  $\sigma_y$, $I$, and 
$\sigma_d$ denote the symmetry transformations necessary to 
calculate $t(\Delta x, -\Delta y;\mu,\nu)$, 
$t(-\Delta x, -\Delta y;\mu,\nu)$, and $t(\Delta y, \Delta x;\mu,\nu)$, 
respectively. Here `$\pm$' means that the corresponding  
hopping is equal to $\pm t(\Delta x, \Delta y; \mu, \nu)$ in the same row,
and `$\pm (\mu',\nu')$' states that the
hopping equals  $\pm t(\Delta x, \Delta y; \mu', \nu')$
in another $(\mu', \nu')$ row.
Note that there is another symmetry relation 
$t(\Delta x, \Delta y; \mu, \nu) = t(-\Delta x, -\Delta y; \nu, \mu)$. 
}
\label{transfer}
\end{table}

\clearpage 

\subsection{\label{results:FeSb2}\label{FeSb2}FeSb$_2$}

\begin{figure}[p]
\centering\includegraphics[height=0.3\columnwidth,angle=0,clip]{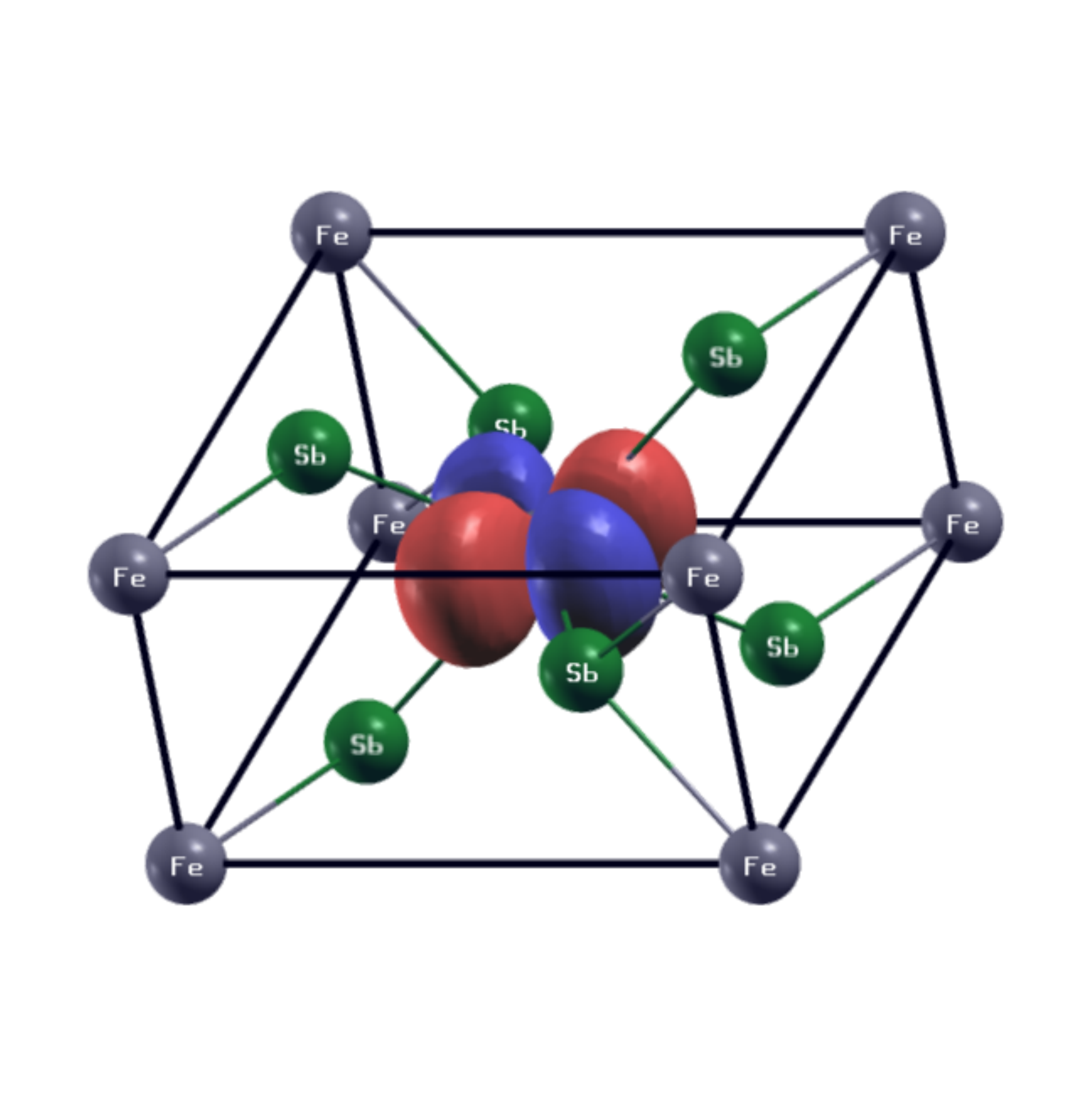}
\includegraphics[height=0.3\columnwidth,angle=0,clip]{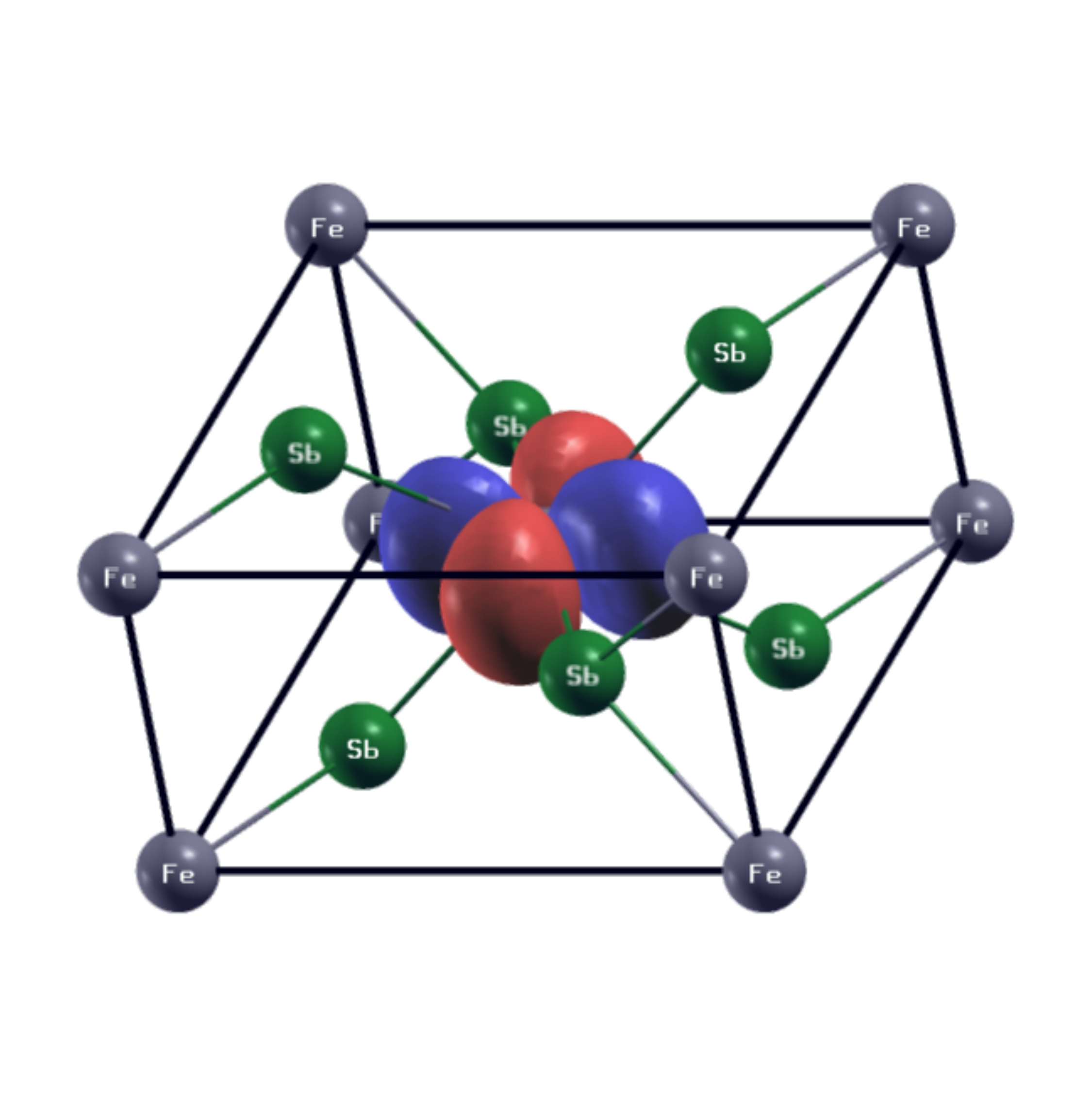}
\includegraphics[height=0.3\columnwidth,angle=0,clip]{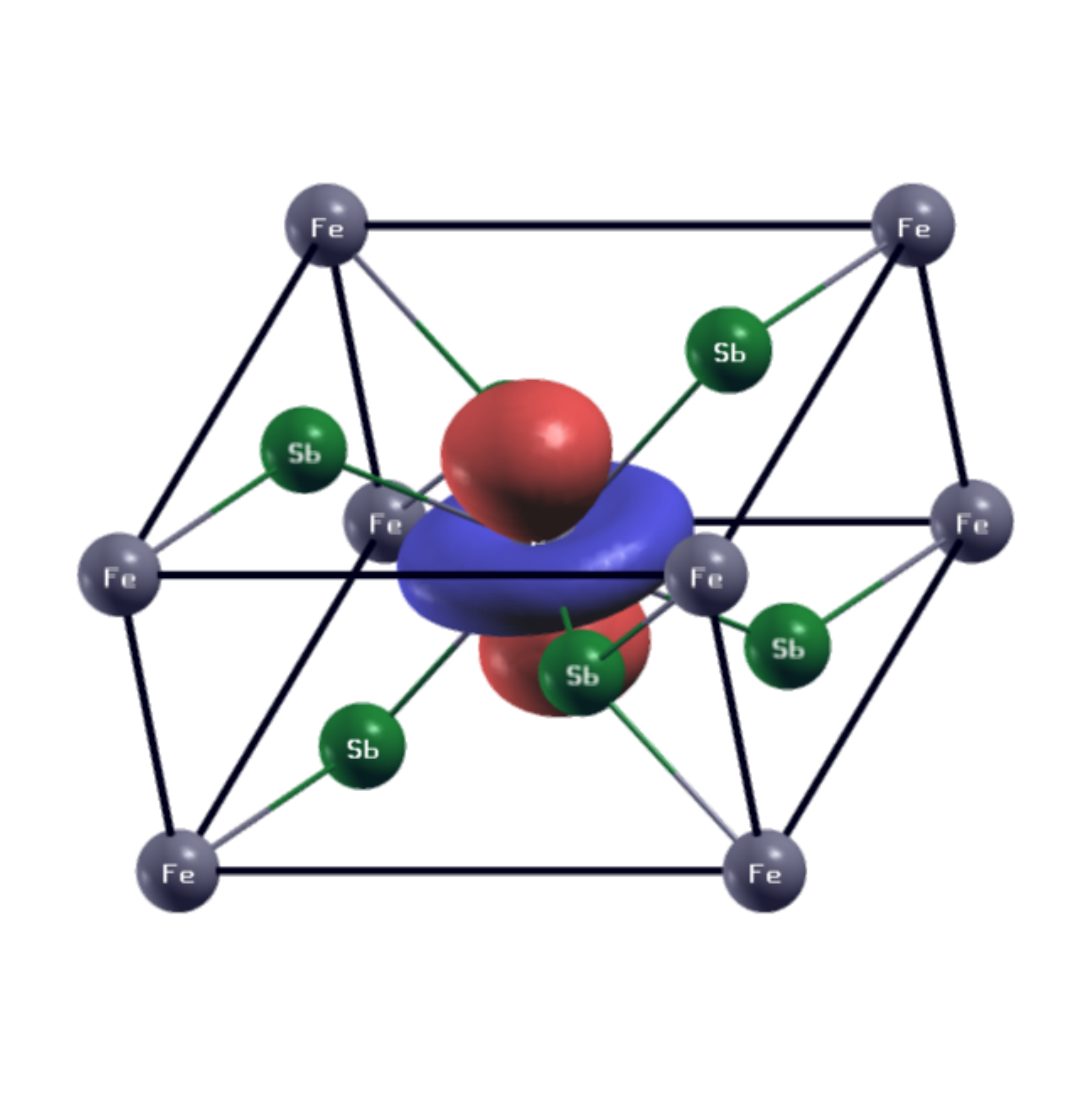}\\
\includegraphics[height=0.3\columnwidth,angle=0,clip]{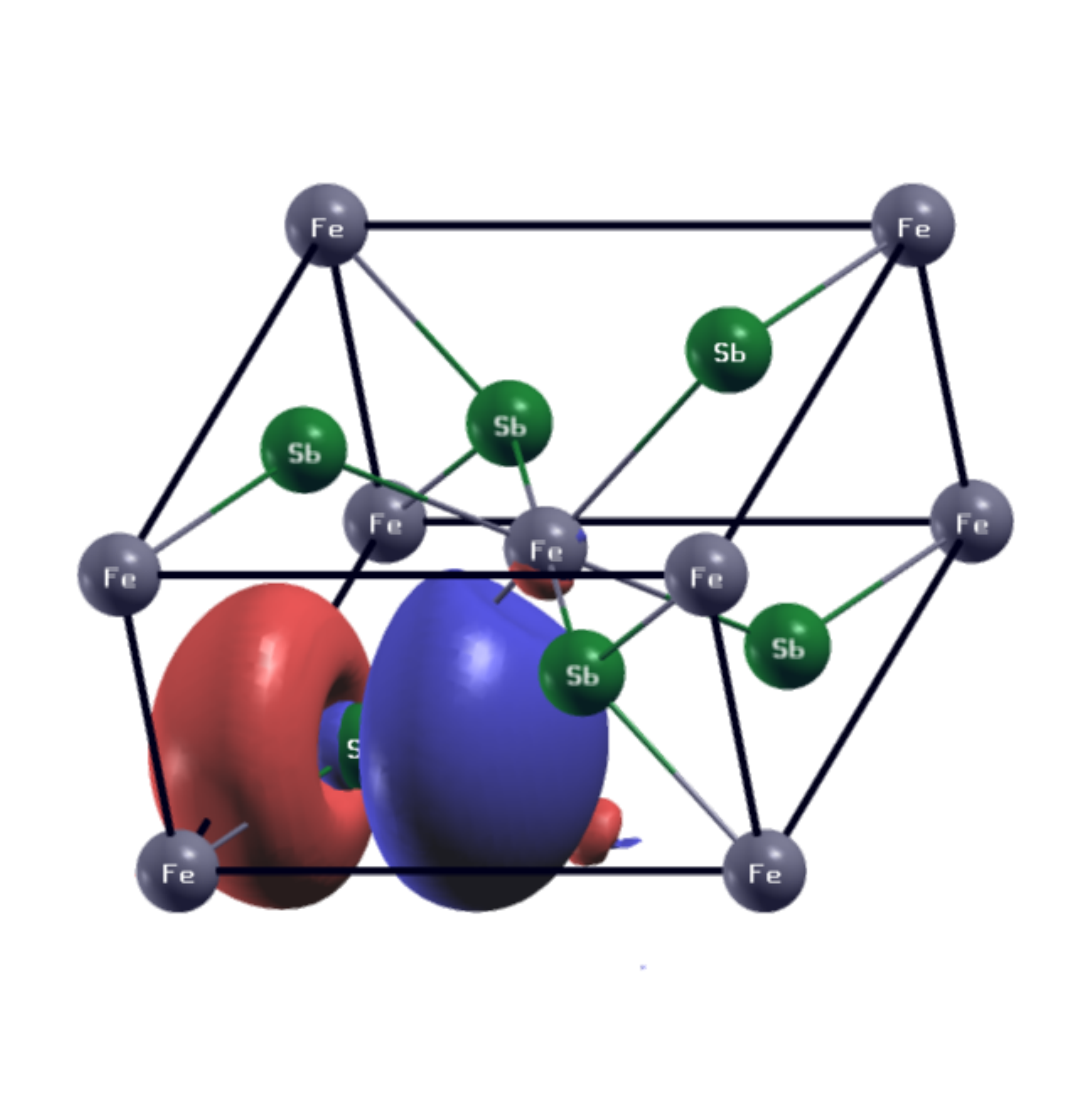}
\includegraphics[height=0.3\columnwidth,angle=0,clip]{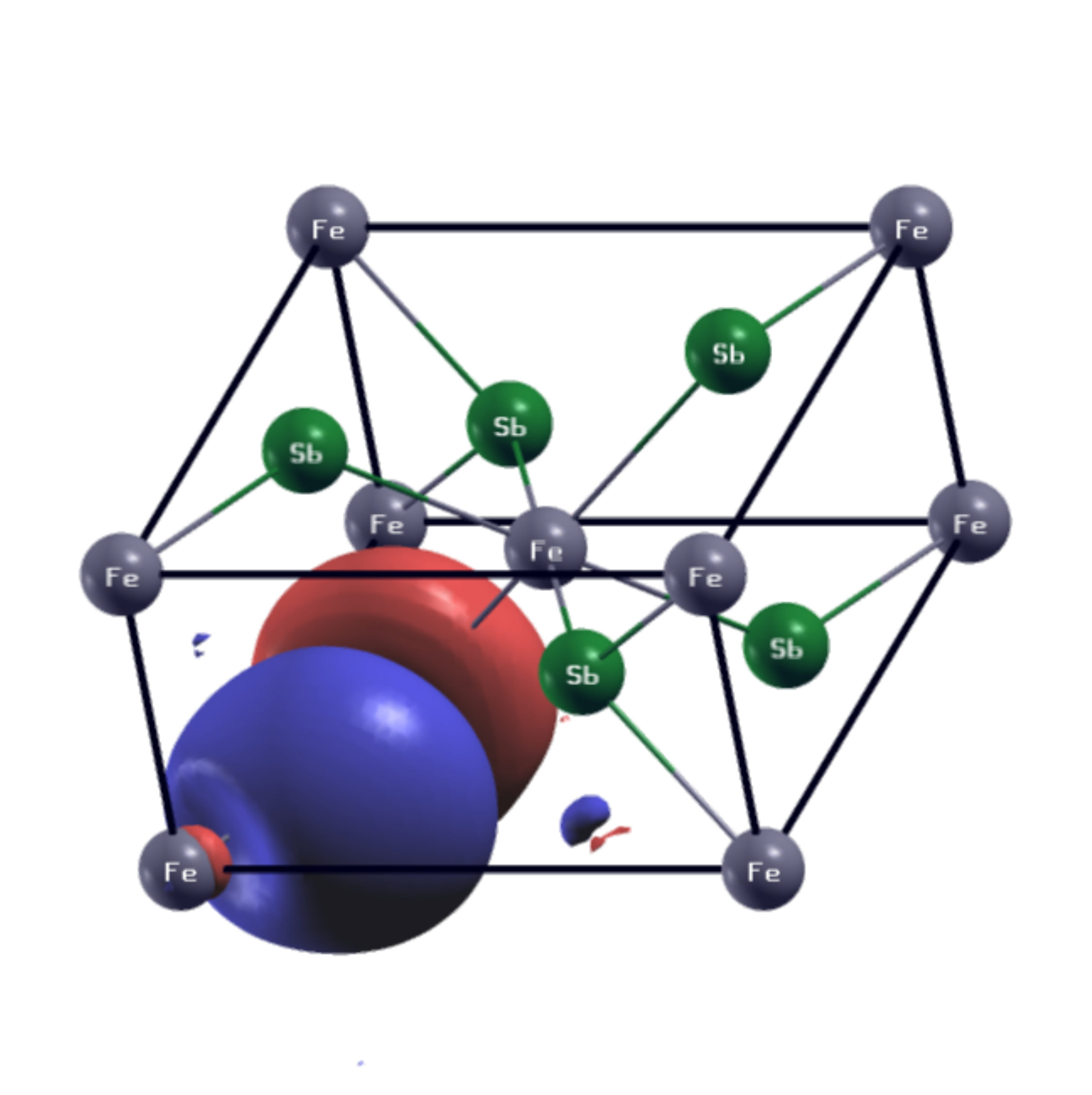}
\includegraphics[height=0.3\columnwidth,angle=0,clip]{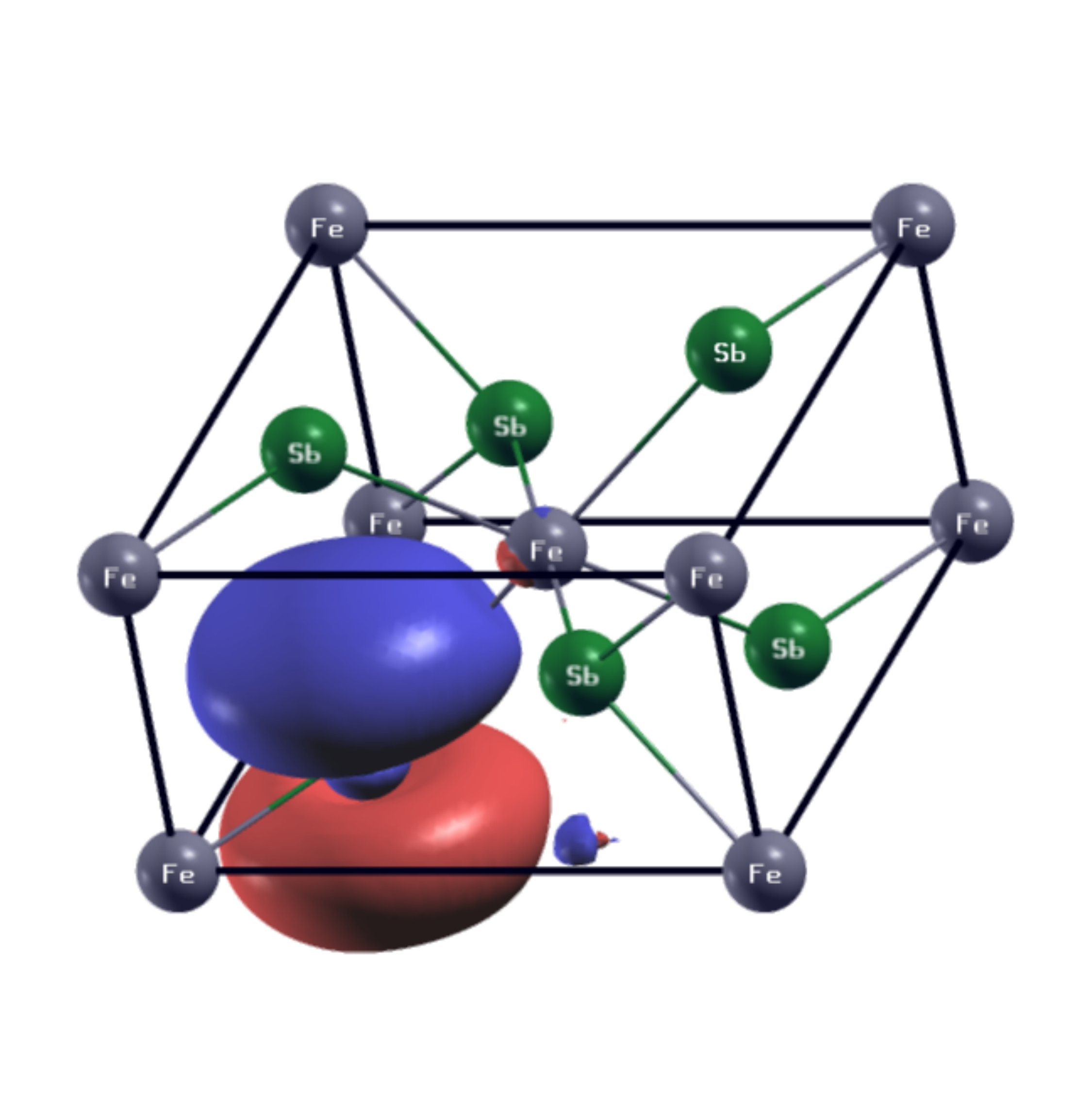}\\
\caption{\label{fig:FeSb2 Wannier orbitals}(color online) The basis of Wannier
orbitals in FeSb$_2$, presented as an isosurface plot in real space within the coordinate system of
the primitive unit cell. In the first row~(from left to right), we show one of the
orbitals with predominant e$_\textrm{g}$ character~(pointing approximately in the
direction of the ligands) and two of the orbitals with mainly t$_\textrm{2g}$
character, respectively. In the second row we show, from left to right, the
p$_\textrm{x}$, p$_\textrm{y}$, p$_\textrm{z}$ orbitals.}
\end{figure}
\begin{figure}[p]
\centering
\psfrag{eV}{\small eV}
\includegraphics[height=0.25\columnwidth,angle=0,clip]{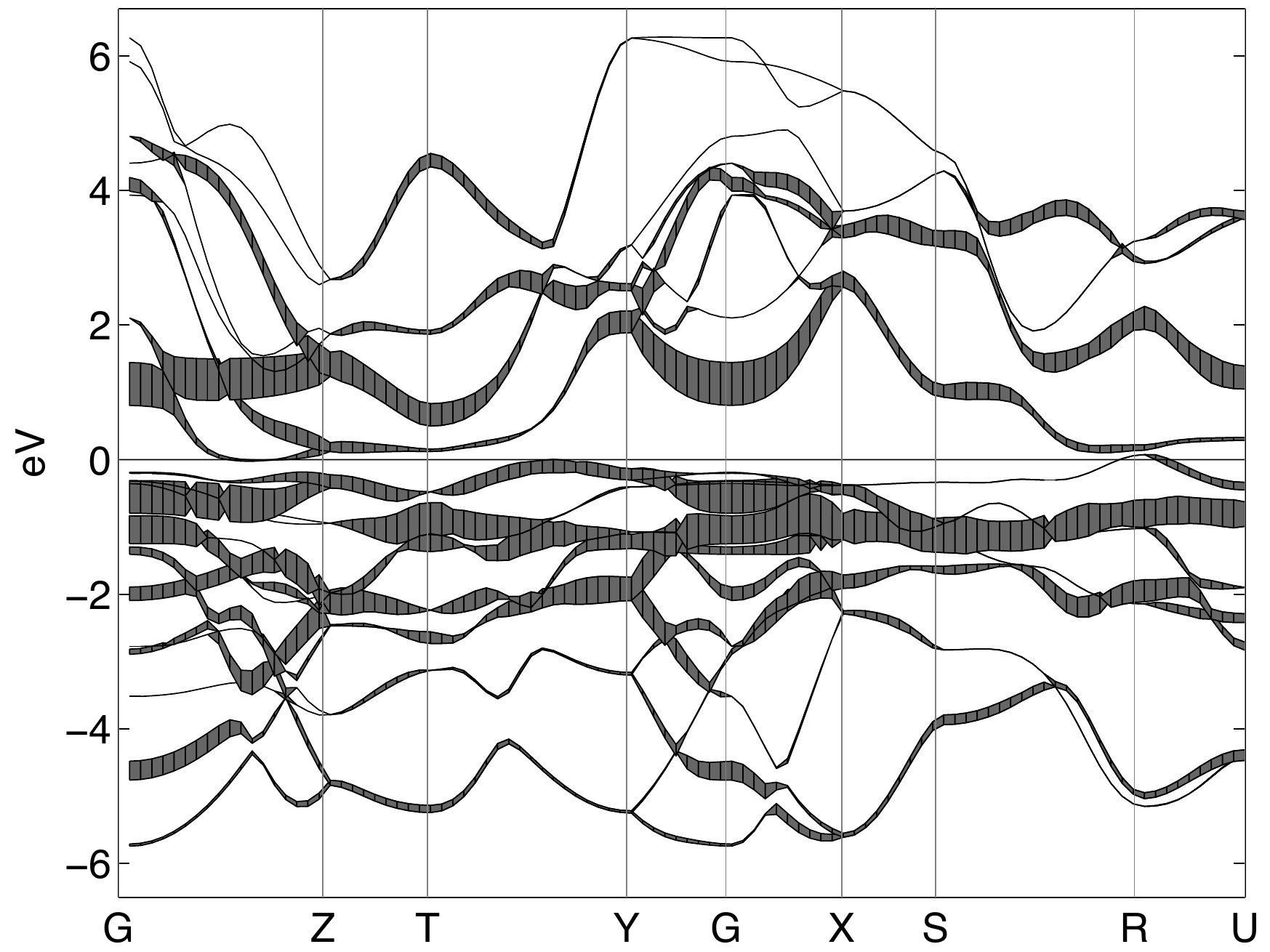}\hspace*{0.1
cm}
\includegraphics[height=0.25\columnwidth,angle=0,clip]{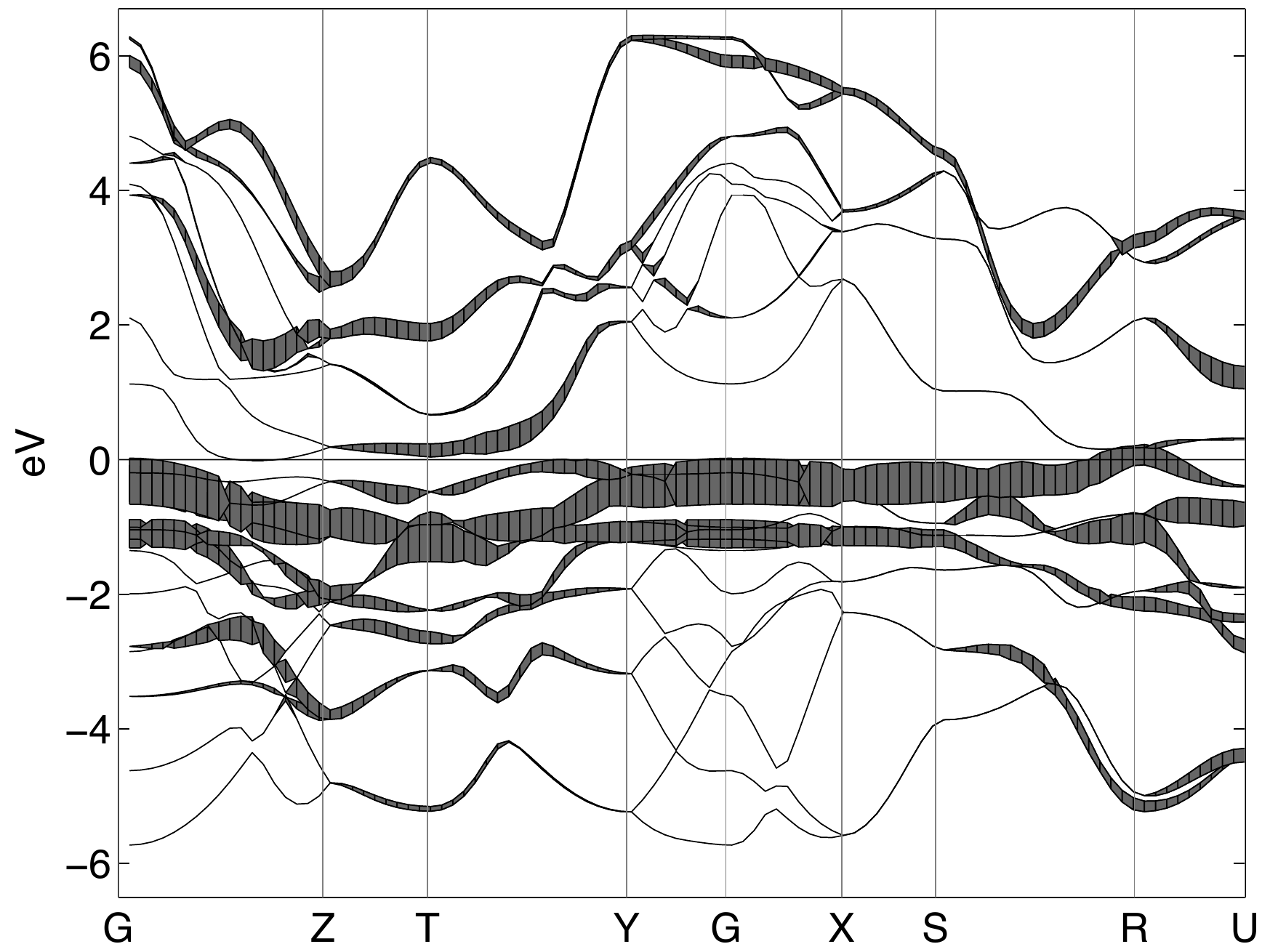}\hspace*{0.1
cm}
\includegraphics[height=0.25\columnwidth,angle=0,clip]{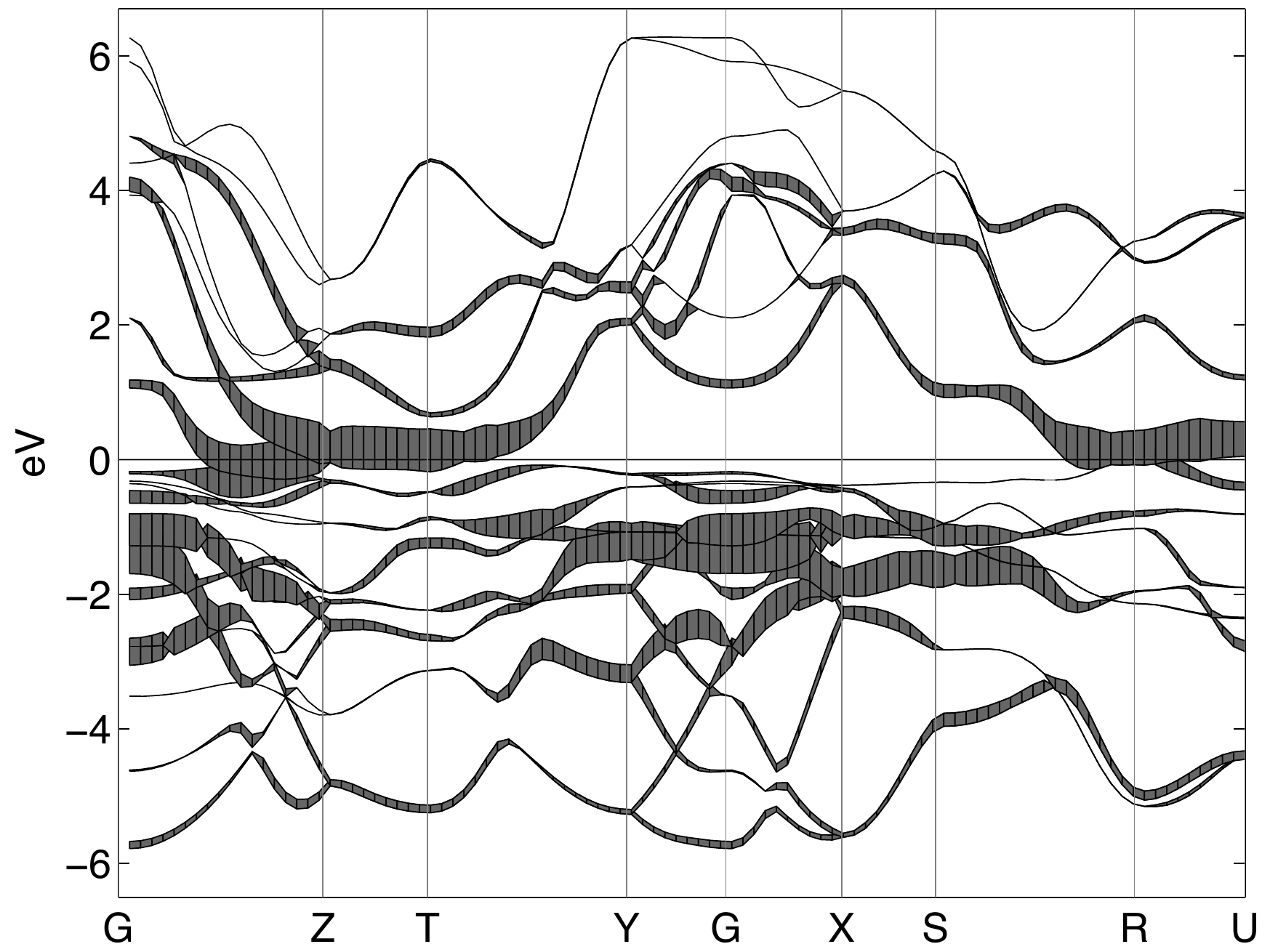}\\
\includegraphics[height=0.25\columnwidth,angle=0,clip]{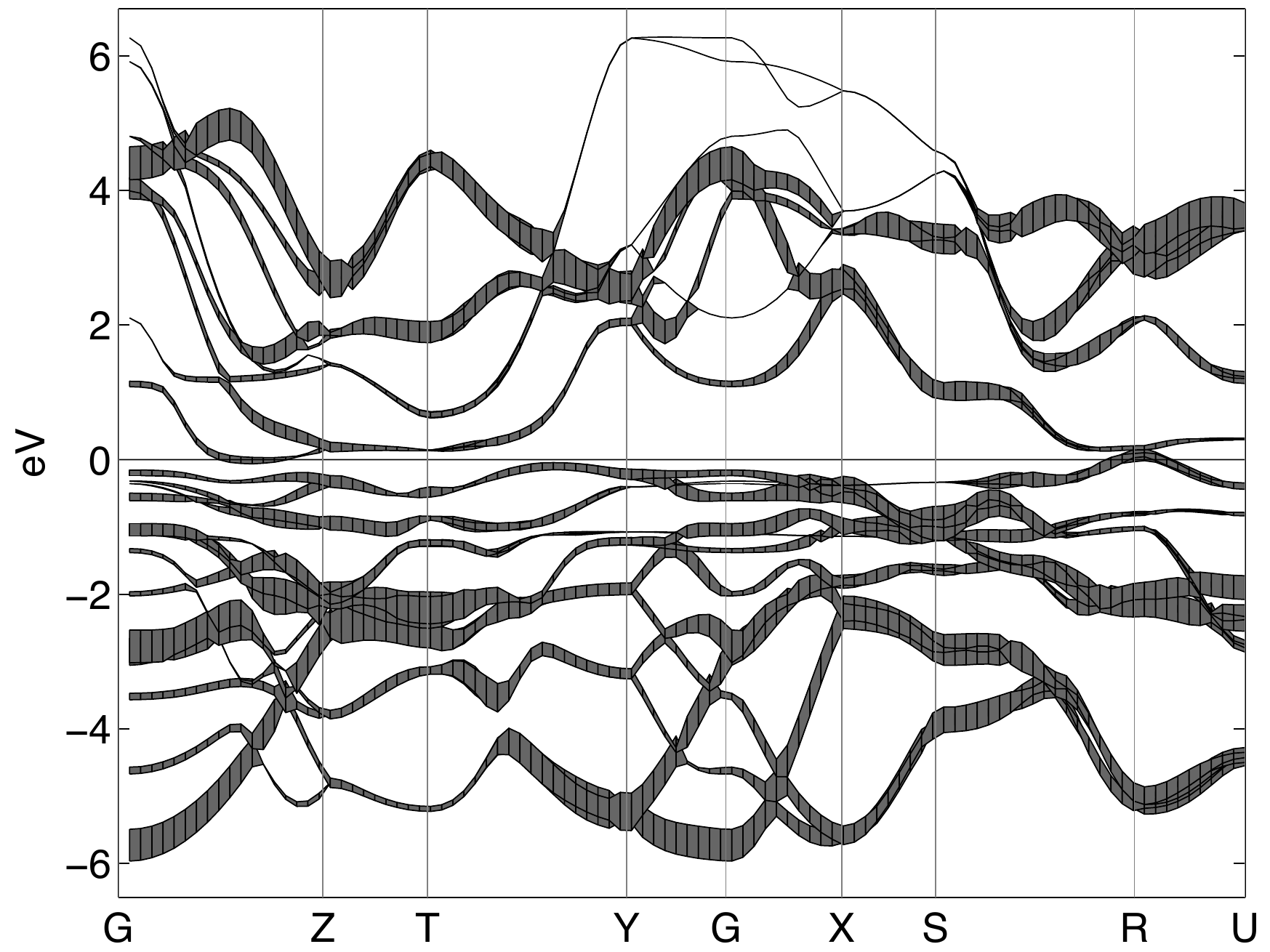}\hspace*{0.1
cm}
\includegraphics[height=0.25\columnwidth,angle=0,clip]{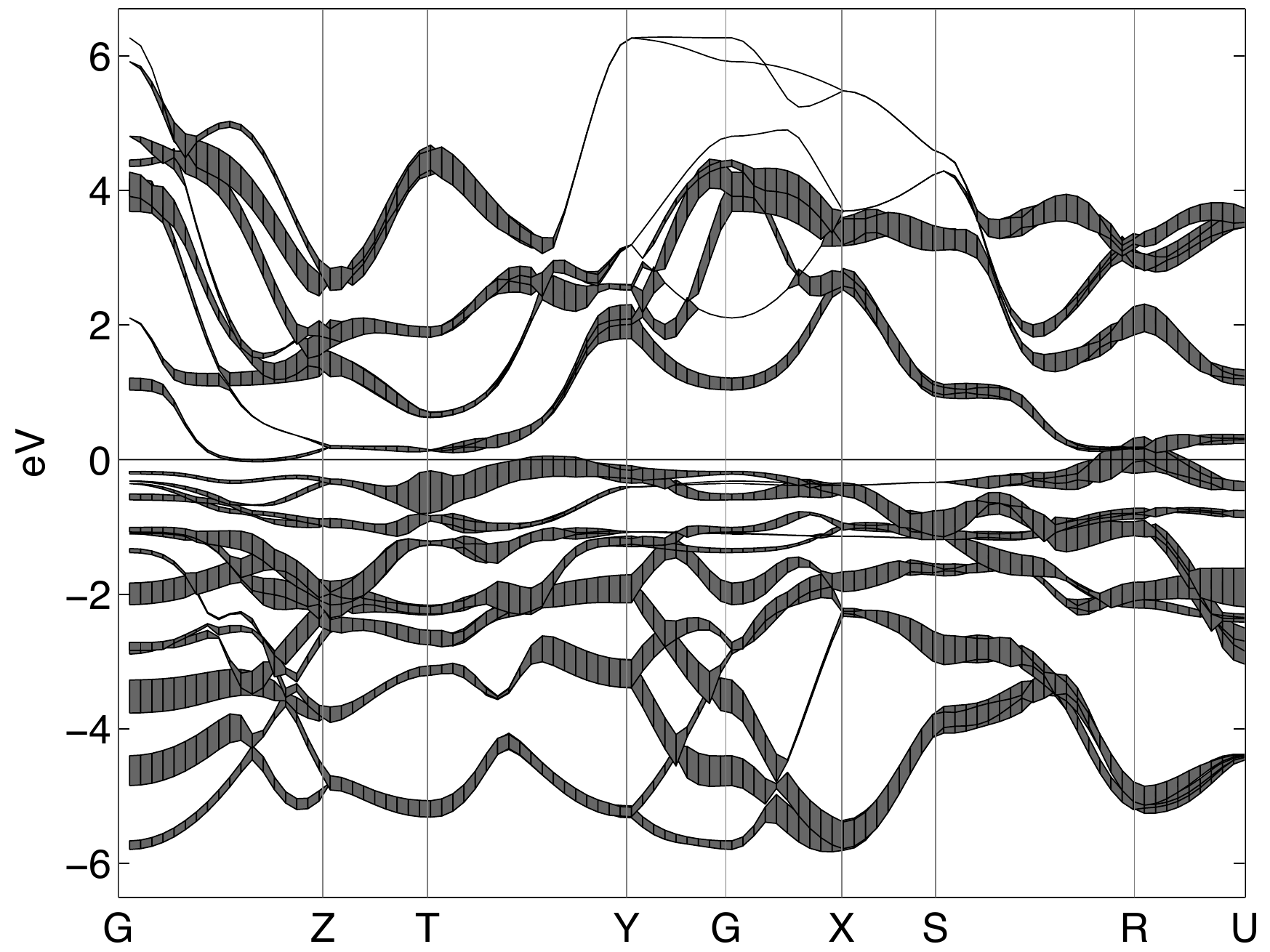}\hspace*{0.1
cm}
\includegraphics[height=0.25\columnwidth,angle=0,clip]{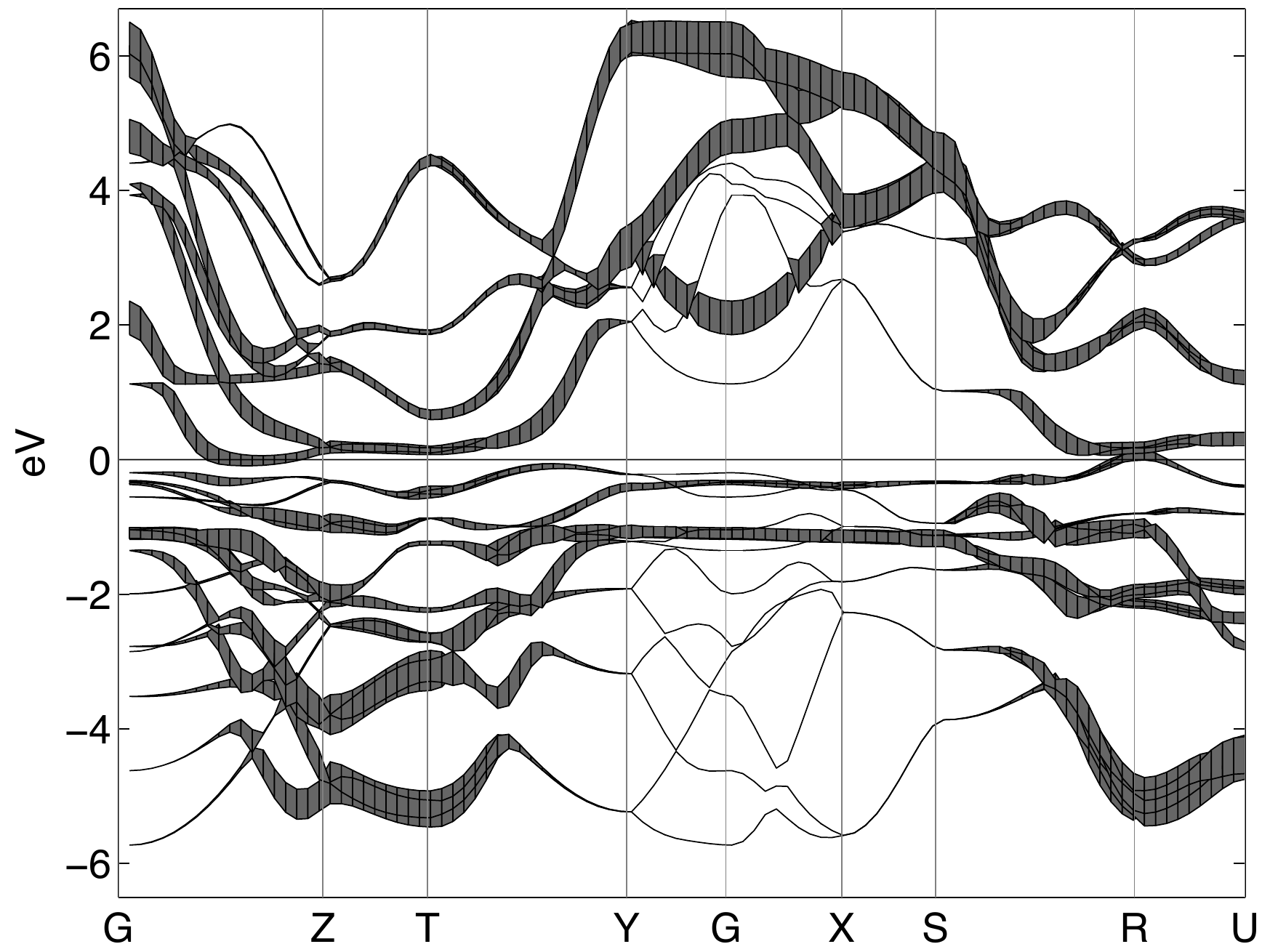}
\caption{\label{fig:FeSb2 fatbands} Partial contributions to the bandstructure
corresponding to the Wannier orbitals visualized in Figure~\ref{fig:FeSb2 Wannier
orbitals}. From top left to bottom right, the widths of the bands indicate the
contribution of the $e_\textrm{g}$, $t_\textrm{2g}$($e_\textrm{g}^{\pi}$),
$t_\textrm{2g}$($a_\textrm{1g}$), $p_\textrm{x}$, $p_\textrm{y}$, and $p_\textrm{z}$ 
Wannier function, respectively. Note the huge $p$-$d$ hybridization in the energy
region~[-4 eV, 4 eV].}
\end{figure}

FeSb$_2$ is a small band gap semiconductor which is intensively
studied \cite{BentienPRB06} for its unusually good thermoelectric properties.  In order
to describe the Coulomb interaction within the Fe-3d
shell and the strong hybridization between the Fe-$3d$ and the Sb-$4p$ electrons on an equal footing,
a Wannier projection onto a $22$ band $pd$ subspace is required.
FeSb$_2$ has an orthorhombic structure belonging to
the space group 58(\emph{Pnnm}), in which
the Fe sites are located at the centers of largely distorted Sb octahedra.
This distortion results in a low point group symmetry at the
Fe site.

The LDA bandstructure, shown (grey solid lines
in Fig.~\ref{fig:FeSb2 fatbands}), is rather
complicated, with about 10 bands in the energy interval of $\pm 2$eV around
 the Fermi
level.  The Wannier functions were constructed from 41 Bloch states  
on a 10x10x10 $\mb{k}-$grid, by projection onto the subspace of 22 MLWF  in the energy region
[-$6$eV,+$6$eV], using the disentanglement procedure of wannier90. 

We performed  Wannier projections with different  frozen
energy windows and found that already the
window [-$6$eV,$3.9$eV] allows for a good description of all the 22 bands in the
whole energy interval of $\pm 6$eV around  the Fermi level.
The corresponding MLWFs are well localized with the spread
ranging from $0.8$\AA$^2$ to $1.2$\AA$^2$ for Fe-$3d$ orbitals and from $3.6$\AA$^2$
to $5.0$\AA$^2$ for the Sb-$4p$ orbitals.  The largest spread of $\simeq 1.2$\AA$^2$\ was obtained for
the two Fe-$e_\textrm{g}$-like Wannier functions which point more towards the direction of
the ligands (see, e.g., first panel in the first row of Fig.~\ref{fig:FeSb2 Wannier
orbitals}), while the other three (Fe-$t_\textrm{2g}$-like) Wannier orbitals are
slightly more localized with a spread of $0.8$\AA$^2\div 1.0$\AA$^2$ (second and
third panel of Fig.~\ref{fig:FeSb2 Wannier orbitals}). The pictures of the three
Sb-$4p$ orbitals are reported with the same scale in the second row of
Fig.~\ref{fig:FeSb2 Wannier orbitals}.
The contributions of the different MLWFs to the bandstructure is shown as
a so-called  fat-band plot in Fig.~\ref{fig:FeSb2 fatbands}, revealing
 a strongly mixed $d$-$p$
character of all bands.

By truncating the Hamiltonian in the MLWF basis set in direct space, we find that
retaining hoppings up to $7.3$\AA\ provides an accurate description
of the LDA bands. A closer analysis shows that the strongest hopping
processes ($1.2 \div 1.8$ eV)  are in the $p$-$p$ sector of the Hamiltonian. 
Relatively large hoppings are found also in the p-d sector (up to $0.6 \div 0.9$ eV) while
the $d-d$ hopping amplitudes reach the maximum value of about $300$meV. 

\section{Summary}
We have presented the implementation of an interface between the FLAPW code Wien2K and 
wannier90 software for the construction of the maximally localized Wannier functions.
The rational of this development is to provide a link between two widely used packages
in the electronic structure community. We tried to provide examples of what we consider
typical application of such a construction such as the construction of
 tight-binding Hamiltonians for complex systems,
unfolding of complicated bandstructures or visualization.
We also considered as an
 applications that have been so far less common
such as strongly spin-orbit coupled Wannier orbitals. Last but not least we point out that
the  overlap matrices [Eq.\ (\ref{eq:mmn})] can find application of its own, e.g. concerning the calculation
of the interaction with an electro-magnetic field 
beyond the dipole approximation.

\section{Acknowledgement}
We would like to thank K. Schwarz, P. Blaha, P. Nov\'ak, K. Nakamura, and T. Miyake
for discussions. J.K. acknowledges the support of the SFB 484 of the Deutsche Forschungsgemeinschaft
and the grant no. P204/10/0284 of the Grant Agency of the Czech Republic.
R.A. was supported by the Next Generation Supercomputing Project of Nanoscience Program
from MEXT, Japan and the Funding Program for World-Leading
Innovative R\&D on Science and Technology(FIRST Program),
K.H. by the EU-Indian cooperation network MONAMI, and P.W. through WK004 of the Austrian Science Foundation (FWF).
J.K. and K.H. thank for hospitality of the Kavli Institute of Theretical Physics and UC Santa Barbara.


\end{document}